\documentclass[journal]{IEEEtran}
\usepackage{cite}
\usepackage[pdftex]{graphicx}
\usepackage[cmex10]{amsmath}
\usepackage{algorithm}
\usepackage{array}
\usepackage[caption=false,font=footnotesize]{subfig}
\usepackage{url}
\usepackage{cases}
\usepackage{bm}
\usepackage{mathtools}
\usepackage{amsmath, amssymb}  
\usepackage{amsmath}
\usepackage{amssymb}
\usepackage{multirow}
\usepackage{float}  
\usepackage{amsthm}
\usepackage{algpseudocode}
\usepackage{indentfirst}
\usepackage{balance} 
\usepackage{enumerate}
\usepackage{txfonts}
\usepackage{setspace} 
\usepackage{epstopdf}
\usepackage{geometry}
\begin{document}
	\title{Eco-friendly Power Cost Minimization for Geo-distributed Data Centers Considering Workload Scheduling}
	\author{\IEEEauthorblockN{Chunlei Sun$^{1,2,3}$, Xiangming Wen$^{1,2,3}$, Zhaoming Lu$^{1,2,3}$, Wenpeng Jing$^{1,2,3}$, Michele Zorzi$^4$, Fellow, IEEE\\}
		\IEEEauthorblockA{
			$^1$Beijing University of Posts and Telecommunications, Beijing, China, 100876\\
			$^2$Beijing Key Laboratory of Network System Architecture and Convergence (BUPT), Beijing, China, 100876\\
			$^3$Beijing Laboratory of Advanced Information Networks (BUPT), Beijing, China, 100876\\
			$^4$University of Padova, Padova, Italy, 35131\\	
			Email: scl1992@bupt.edu.cn}
}
	\maketitle
	\begin{abstract}
		
		The rapid development of renewable energy in the energy Internet is expected to alleviate the increasingly severe power problem in data centers, such as the huge power costs and pollution. This paper focuses on the eco-friendly power cost minimization for geo-distributed data centers supplied by multi-source power, where the geographical scheduling of workload and temporal scheduling of batteries' charging and discharging are both considered. Especially, we innovatively propose the Pollution Index Function to model the pollution of different kinds of power, which can encourage the use of cleaner power and improve power savings. We first formulate the eco-friendly power cost minimization problem as a multi-objective and mixed-integer programming problem, and then simplify it as a single-objective problem with integer constraints. Secondly, we propose a Sequential Convex Programming (SCP) algorithm to find the globally optimal non-integer solution of the simplified problem, which is non-convex, and then propose a low-complexity searching method to seek for the quasi-optimal mixed-integer solution of it. Finally, simulation results reveal that our method can improve the clean energy usage up to 50\%--60\% and achieve power cost savings up to 10\%--30\%, as well as reduce the delay of requests.

	\end{abstract}
	\begin{IEEEkeywords}
		Geo-distributed data centers, eco-friendly power cost minimization, pollution index, multi-source power, workload scheduling		
	\end{IEEEkeywords}
	\section{Introduction}	

Over the last few years, the demand for cloud computing has grown rapidly, and more and more data and computation services are migrated to geo-distributed data centers. Although this has brought significant benefits, the problem of power consumption is increasingly serious with the growth of data centers. According to \cite{1GoogleDC_online}, the total number of servers in the geo-distributed data centers of Google, Microsoft and Akamai were almost 1 million, 200,000 and 70,000, respectively, and their corresponding power costs were on the order of millions of dollars per year. In 2013 the electricity consumed by data centers across the US was up to 91 million MW.h, and will account for 20\% of the national annual power consumption by 2030 according to some predictions\cite{2DCcost_online}. Furthermore, it is estimated that just a 1-MW data center powered by thermal power can cause over 10,000 metric tons of $\rm{CO}_2$ emissions annually \cite{Li2013Chameleon}. Therefore, it is important to improve the usage of clean energy as well as to reduce the power cost of data centers. 

The recent development of the energy Internet offers further opportunities for eco-friendly power cost optimization of data centers. The concept of energy Internet was first proposed by Jeremy Rifkin in 2008 \cite{Rifkin2012The}. Since 2010, the energy Internet has gradually gained worldwide interests, and a series of national projects have been launched, such as the Future Renewable Electric Energy Delivery and Management system (FREEDM) in the US \cite{FREEDM_Report}, the E-Energy program in Germany \cite{BDI_E-energy} and the global energy Internet project in China\cite{Global_EI}.  
The effects of the energy Internet on the power use of geo-distributed data centers can be summarized in the following four aspects:
\begin{enumerate} 
	\item The implementation of smart electricity price, varying in different times and areas, can encourage the economical workload scheduling among geo-distributed data centers to reduce the total power cost. 
	
	\item The development of storage technology allows power to be generated and consumed at different times, which is important for the supply-demand balancing and peak load shifting. 
	\item The development of renewable energy can improve the use of clean energy, and reduce carbon emissions.
	\item The development of electricity-sale companies can break the monopoly of the traditional electricity market, and can make consumers buy power from multiple sources at lower price. 
\end{enumerate}

In this paper, we consider the scenario of multi-source power supply in the energy Internet, and aim at the eco-friendly power cost minimization for geo-distributed data centers by the combination of $(1)$ green and economical strategies of buying power form different sources, $(2)$ geographical scheduling of workload, and $(3)$ temporal scheduling of batteries' charging and discharging, or called power storage.

\subsection{Related Work}
In the literature, many studies have been done about power-cost-driven workload scheduling in geo-distributed data centers. For instance, \cite{survey2015} is a survey about the energy-aware geographical workload balancing, and \cite{MapReduce2014} and \cite{Mashayekhy2015Energy} proposed some energy-aware scheduling methods towards MapReduce jobs in data centers. \cite{Rao2012Coordinated} \cite{Shao2014Optimal} \cite{Luo2015Spatio} and \cite{Luo2014Temporal} proposed a series of linear programming based methods, while \cite{Yao2012Data} and \cite{Yao2014Power} proposed a stochastic process based method to minimize the power cost, where more workloads were distributed to the data centers with lower local electricity price and more delay-tolerant requests were scheduled to the time slots with cheaper electricity. Besides the workload scheduling, the temporal scheduling of power storage was taken into consideration in \cite{Yu2015Joint}, and the cooperation between the power use of data centers and the power storage of electric vehicles was studied in \cite{Li2014Integrated} \cite{Yu2016Joint} and \cite{Yu2016Distributed}. These strategies can greatly cut down the electricity bills of data centers, but do not consider the use of renewable energy or the scenario of multi-source power. 

Other works have researched the use of renewable energy and multi-source power for data centers. The opportunities and challenges of harnessing renewable energy for data centers were well surveyed in \cite{Deng2014Harnessing}. Besides, \cite{Deng2013SmartDPSS} and \cite{Deng2013MultiGreen} proposed a stochastic process based method for a single data center to buy power from multiple sources, while \cite{Liu2016Towards} formulated a machine learning based method and designed a multi-source power distribution architecture. These works assumed that data centers could buy power from the power grid, but could also obtain power from their own storage devices and renewable energy generators. However, these kinds of researches rarely consider the cooperation of multiple data centers.

Several works have researched the power cost minimization for geo-distributed data centers by jointly considering renewable energy, multi-source power and workload scheduling. \cite{Islam2017A} proposed a bid mechanism to reduce the carbon footprint. \cite{Zhou2013Carbon} and \cite{Toosi2017Renewable} proposed some strategies of geographical workload balancing, where more workloads were dispatched to the areas with lower carbon power or more renewable power. These strategies mainly concentrated on the workload scheduling, while not considering power buying or harvesting. \cite{Yu2014Carbon} and \cite{Yu2015Energy} proposed a carbon-aware power cost minimization strategy, where it was assumed that data centers could harvest power from their private green micro-grid and more workload would be distributed to data centers equipped with larger micro-grids to reduce power cost and carbon emission. However, the cost of building a private micro-grid is very high and many operators usually cannot afford it. Different from this, \cite{Rakesh2016Optim} and \cite{Rakesh2017Cost} assumed that data centers could directly buy clean power generated by different kinds of renewable energy from the market, but they did not consider the differences of pollution among various renewable energy and forced operators to buy a fixed minimum percentage of clean power, which is not very flexible. In this paper, we aim to propose a green and economical strategy to buy power from multiple sources dynamically in terms of their pollution and real-time price. The main contributions of this paper are introduced below.

	Besides, Liang Yu et al. proposed a carbon-aware load balancing strategy for geo-distributed data centers in \cite{Yu2014Carbon} and then further consider the uncertain power outage in \cite{Yu2015Energy}. They assumed that cloud service operators would be interested to build private micro-grid with the objective of protecting data centers from power outages caused by bulk power grid and cutting down their power bills. In their proposed strategy, more workload would be balanced to data centers equipped with larger micro-grid to reduce power cost and carbon emission. But, in fact, the cost of building private micro-grid is extremely immense and many operators usually cannot afford it. Different from this, 

 \subsection{Our Contributions}
  The main objective of this paper is to make eco-friendly power cost minimization for geo-distributed data centers by combining clean power buying, temporal power-storage scheduling and geographical workload scheduling, which is denoted as Problem P\&W. In particular, we model the total power cost as the weighted sum of the pollution cost and the monetary cost, and innovatively propose the new Pollution Index Function (PIF) to model the pollution costs of different kinds of power, which can efficiently increase the use of cleaner power. We firstly formulate Problem P\&W as a  multi-objective programming problem with integer constraints, and then simplify it into a single-objective problem under the assumption that the transmission delay will not exceed an upper bound. Secondly, we propose an SCP algorithm to find the globally optimal non-integer solution of the simplified version of Problem P\&W, and propose a low-complexity searching method to seek for the quasi-optimal mixed-integer solution of the simplified Problem P\&W. Finally, we give the condition when Problem P\&W is equal to its simplified version, and Problem P\&W is solved eventually. 
  
  Our main contributions can be summarized in three aspects as below.  
  	\begin{enumerate} 
  	  	\item We formulate the eco-friendly power cost minimization for geo-distributed data centers as Problem P\&W, which can improve the clean energy usage up to 50\%--60\% and achieve power cost savings up to 10\%--30\%, as well as reduce the delay of requests.  	  	

  		\item We propose the new PIF to model the pollution costs of different kinds of power, by which our proposed strategies can encourage the power buying from multiple sources and improve the usage of clean energy.
  		\item We propose an efficient SCP algorithm. Mathematical proofs show that it can obtain the globally optimal non-integer solution of the simplified Problem P\&W, although it is non-convex.       		

  	\end{enumerate}
 
 The rest of this paper is organized as follows. Firstly, we introduce the system architecture and formulate the model of this power cost minimization problem in Section II. Secondly, we propose the related algorithm to solve this problem in Section III. Thirdly, we give simulation results about the performance of our model in Section IV. Finally, we conclude the paper in Section V.
  
	\section{Modeling and Formulation }		
	In this section, we firstly describe the system model, where we consider a geo-distributed data center system with multi-source power supply. Then, we model the costs of eco-friendly power buying, storage scheduling and workload scheduling. The eco-friendly power cost minimization problem is finally formulated as a multi-objective programming problem.
	\subsection{Overview of the System Model}\normalsize
	A cloud computing system usually contains several portal servers and a group of geo-distributed back-end data centers. Each portal server aggregates user requests originating from its service area and then dispatches the corresponding service tasks to back-end data centers, which possess massive computing, memory and data resources. Our system model is given as follows.
	
\begin{figure}[htbp]
		\centering
		\includegraphics[width=0.5\textwidth]{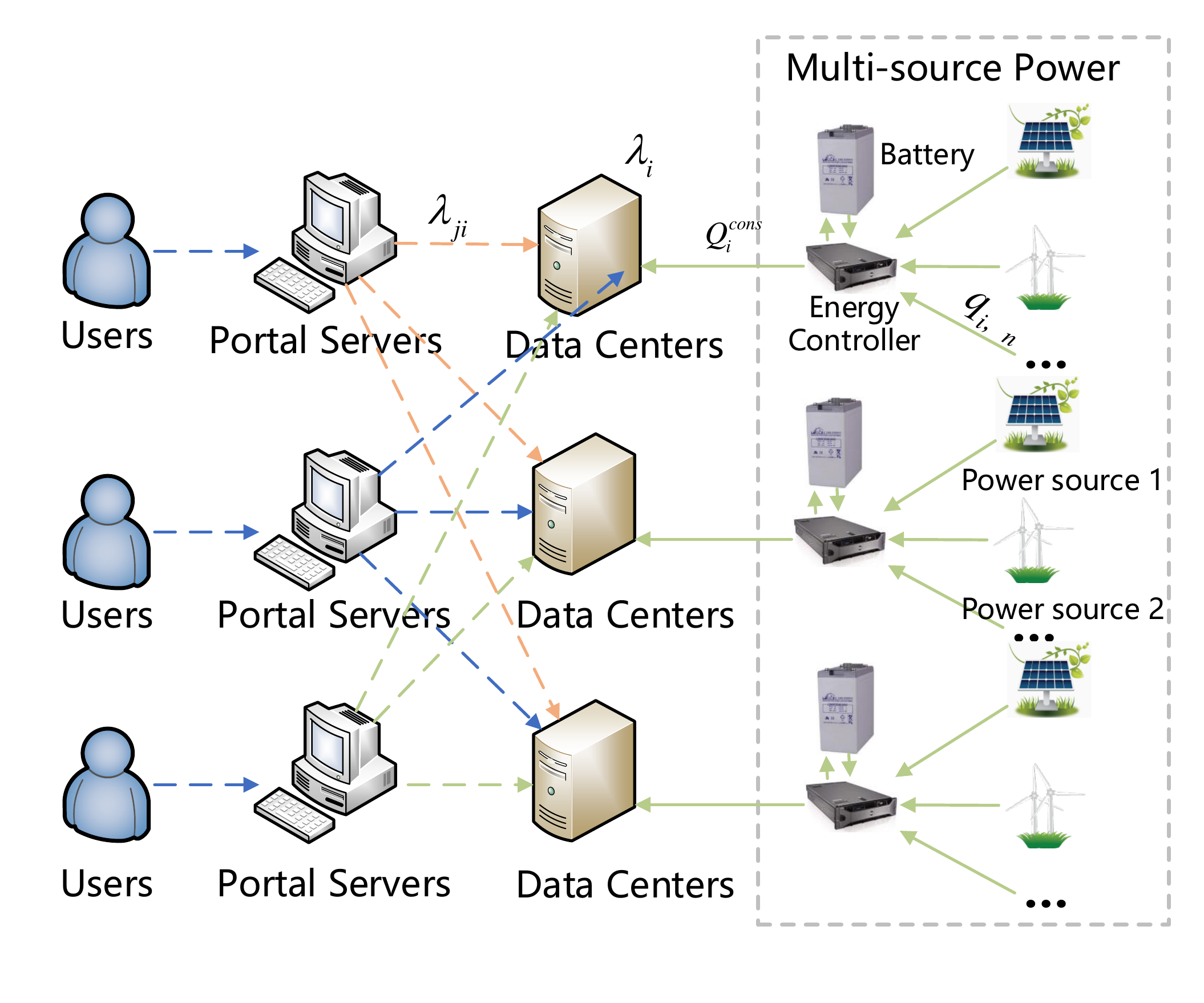}
		\caption{System model of geo-distributed data centers}
		\label{SystemModel}
	\end{figure}

	As shown in Fig.\ref{SystemModel}, there are $J$ portal servers $PS_j$ and $I$ geographically separate data centers $DC_i$ in the cloud computing system, where $j\in\mathcal{J}=\{1,...,J\}$ and $i\in\mathcal{I}=\{1,...,I\}$. We regard this cloud computing system as a discrete time system and discuss the optimization problem in one time slot. We assume that the workload arrival rate to the $j$th portal server is $\lambda_j^{PS}$ per second, and $L=\sum_{j=1}^{J}\lambda_j^{PS}$ is the total request rate of the whole system, where $0\leq L\leq L_{max}$. Let $\lambda_{ji}$ define the request rate from the $j$th portal server to the $i$th data center, so that $\sum_{i=1}^{I}\lambda_{ji}=\lambda_j^{PS}$ and $\sum_{j=1}^{J}\lambda_{ji}=\lambda_i^{DC}$, where $\lambda_i^{DC}$ is the total request rate to the $i$th data center. 
	
	In addition, we define $M_i$ and $m_i^{ac}\leq M_i$ as the number of servers and active servers in the $i$th data center, respectively. According to \cite{Cao2013Optimal}, the power consumption of the $i$th data center in the present time slot can be approximately calculated as
	\begin{equation}
		\label{eq1}
		Q_i^{cons}=\tau^t\times(m_i^{ac}s_i^{\alpha}+\beta_i) \, , \quad m_i^{ac}\in\{1,2,...,M_i\} \, ,
	\end{equation}
	where $\tau^t$ is the length of the present time slot measured in
	hours, $s_i^{\alpha}$ is a parameter related to the CPU frequency, and $\beta_i$ represents the basic power consumption of idle servers, e.g., due to the air-conditioning system, communication devices, etc. We note that $m_i^{ac}$ must be an integer, and $m_i^{ac}\geq 1$ means that at least one server is active in each data center. When $m_i^{ac}=M_i$, all servers in the $i$th data center are active and the maximum
	power consumption $P_i^{max}$ of $DC_i$ will be reached.
	
	Furthermore, we consider that each data center is equipped with a smart energy controller to realize the joint control of buying power, charging or discharging batteries, and supplying power for loads. In a deregulated electricity market, data centers can buy power from more than one source according to the price, pollution, or other factors. We assume that the $i$th data center can buy $N_i$ kinds of power and denote $q_{i,n}$ as the amount of $n$th power bought by the $i$th data center, where $q_{i,n}\geq0,\forall n\in\mathcal{N}_i=\{1,...,N_i\}$. Besides, we let $Q_i^{bat}>0$ denote the power amount actually used when we charge batteries, let $Q_i^{bat}<0$ denote the power amount we actually get when batteries discharge, and let $Q_i^{bat}=0$ when no power is charged or discharged.  
	Then the total power that the $i$th data center needs to buy can be calculated as
	\begin{equation}
	\label{supply-demand}
	Q_i^{buy}=Q_i^{cons}+Q_i^{bat}=\sum\limits_{n=1}^{N_i}q_{i,n} \, .
	\end{equation}
	Here, we assume that the amount of power discharged by storage devices is always less than that used by servers in the $i$th data center, so that $Q_i^{buy}\geq 0$. 
	\subsection{Power Cost model with Pollution Index Function}\normalsize 
	In this subsection, we will model the power cost of data centers when we buy power from multiple sources in one time slot, and define the total power cost as the weighted sum of the pollution cost and the monetary cost.
	
	Firstly, the Pollution Index Function (PIF) is proposed to model the pollution cost\footnote{The Pollution cost can be regarded as a kind of virtual cost or real cost according to our demand. In this paper, it needs to be paid as a real cost.} of different power sources. We denote $f^{PI}(q_{i,n})$ as the PIF with respect to $q_{i,n}\geq 0,\forall n\in\mathcal{N}_i$. In order to encourage eco-friendly power consumption, the following three assumptions are required to be met for PIF. 
	\begin{enumerate}
		\item $f^{PI}(q_{i,n})$ is larger for more polluting power sources, for a given $q_{i,n}$. 
		\item $f^{PI}(q_{i,n})$ is a strictly increasing function of $q_{i,n}$, and $f^{PI}(0)=0$.  
		\item $f^{PI}(q_{i,n})$ is a strictly convex function of $q_{i,n}$. 
	\end{enumerate}
According to the first assumption, cleaner power always leads to less pollution cost, which will encourage the use of cleaner energy. According to the remaining two assumptions, the total cost, the unit cost and the marginal cost \cite{Samadi2010Optimal} of electricity will all increase with the growth of $q_{i,n}$, which will penalize the waste of power and encourage users to save power. A typical example of PIF that meets the above three assumptions is the quadratic function
	\begin{equation}
		f^{PI}_{i,n}(q_{i,n})=a_{i,n}q_{i,n}^2,\quad q_{i,n}\geq 0,\, a_{i,n}> 0, \, \forall i\in \mathcal{I},\, \forall n\in \mathcal{N}_i ,
	\end{equation}
	where $a_{i,n}$ is a coefficient. However, we should emphasize that the characteristic that the marginal cost will increase with the growth of $q_{i,n}$ may be unfair to the users who always need a huge amount of power, although it can contribute to the power savings. For example, a 1-MW data center needs much more power than a 1kW data center in general, and needs to pay a higher unit cost, which further increases its huge power bills. Thus, we define $a_{i,n}=\frac{\gamma_{i,n}}{\tau^tP_i^{max}}$ where $\gamma_{i,n}$ is the pollution factor of power supplied by the $n$th supplier located in the area of $DC_i$, and $\tau^tP_i^{max}$ is the maximum amount of power that can be used by $DC_i$ in one time slot. The introduction of divisor $\tau^tP_i^{max}$ makes a 1-MW data center using 1 MW.h and a 1-kW data center using 1 kW.h be penalized by the same degree. And we denote $f^P=\tau^tP_i^{max}$ as the power factor.
	 
	Secondly, let $p_{i,n}$ denote the electricity price of the $n$th power source for the $i$th data center, so that the corresponding monetary cost is $p_{i,n}q_{i,n}$. 
	
	Finally, we define the total power cost for the $i$th data center as the weighted sum of the pollution cost and the monetary cost as follows:  
	\begin{equation}\label{q-n}
		\begin{aligned}
			& F_i^{PC}(\mathbf{q}_{i})=\sum\limits_{n=1}^{N_i}a_{i,n}q_{i,n}^2+p_{i,n}q_{i,n}\\ 
			&\begin{array}{r@{\quad}r@{}l@{\quad}l}
				{\rm{s.}}\  {\rm{t}}{\rm{.}}& \sum\limits_{n=1}^{N_i}q_{i,n}=Q_i^{cons}+Q_i^{bat}\\
				&\quad  q_{i,n} \geq 0, \forall n \in \mathcal{N}_i \, ,
			\end{array}
		\end{aligned}
	\end{equation} 
where $\mathbf{q}_i=\{q_{i,1},\cdots,q_{i,N_i}\}$. In addition, we denote $v_i$ and $\bar{v}_i$ as the marginal cost and unit cost of power, which are given by
\begin{equation}
\begin{aligned}
& v_i=\sum\limits_{n=1}^N\frac{\partial F_i^{PC}}{\partial q_{i,n}}\\
&\bar{v}_i=
	\left\{ 
	\begin{array}{cl}  
	\frac{F^{PC}_i(\mathbf{q}_{i})}{\sum_{n=1}^{N_i}q_{i,n}}\, , \quad \sum_{n=1}^{N_i}q_{i,n}> 0\\
	\quad 0\, , \qquad \, \sum_{n=1}^{N_i}q_{i,n}= 0
	\end{array}\quad .
	\right.
\end{aligned}
\end{equation} 

	\subsection{Model of Power-Storage Scheduling}\normalsize
In this subsection, we consider both the energy conversion efficiency and the potential cost to model the cost of power-storage scheduling. We denote $C_i$ as the capacity of batteries in the $i$th data center, $c_i$ as the initial amount of electricity stored in the batteries, $\Delta_i$ as the power actually charged in the batteries or the power\footnote{We are not interested in the actual type of energy stored in the batteries and are only interested in how much power it can convert to, so that $C_i$, $c_i$ and $\Delta_i$ are all denoted in terms of the amount of power.} actually used when the batteries discharge, and $\delta_i=\frac{\Delta_i}{\tau^t C_i}$. When $\Delta_i\geq 0$, batteries charge. Otherwise, batteries discharge.

\begin{figure}[!htpb]
			\centering
			\subfloat[An example of $\eta_i$]{\includegraphics[width=4.3cm]{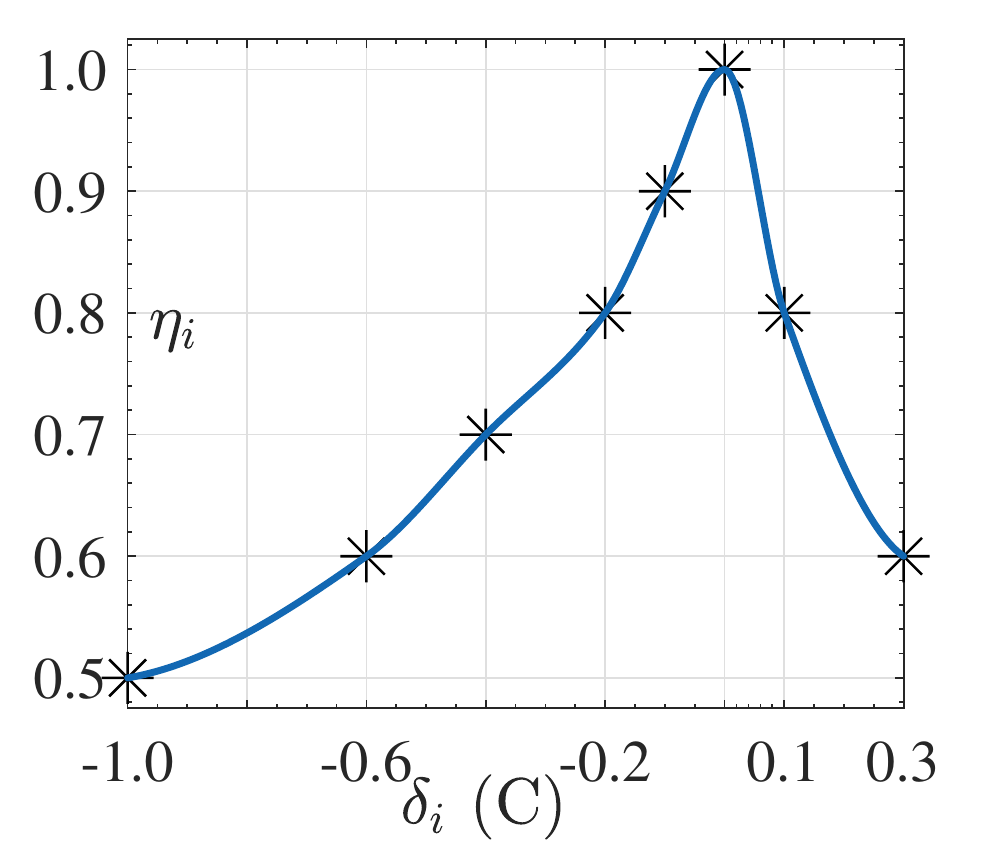}
				\label{nihe1}} 
			\hfill
			\subfloat[An example of $\eta^{'}_i$]{\includegraphics[width=4.3cm]{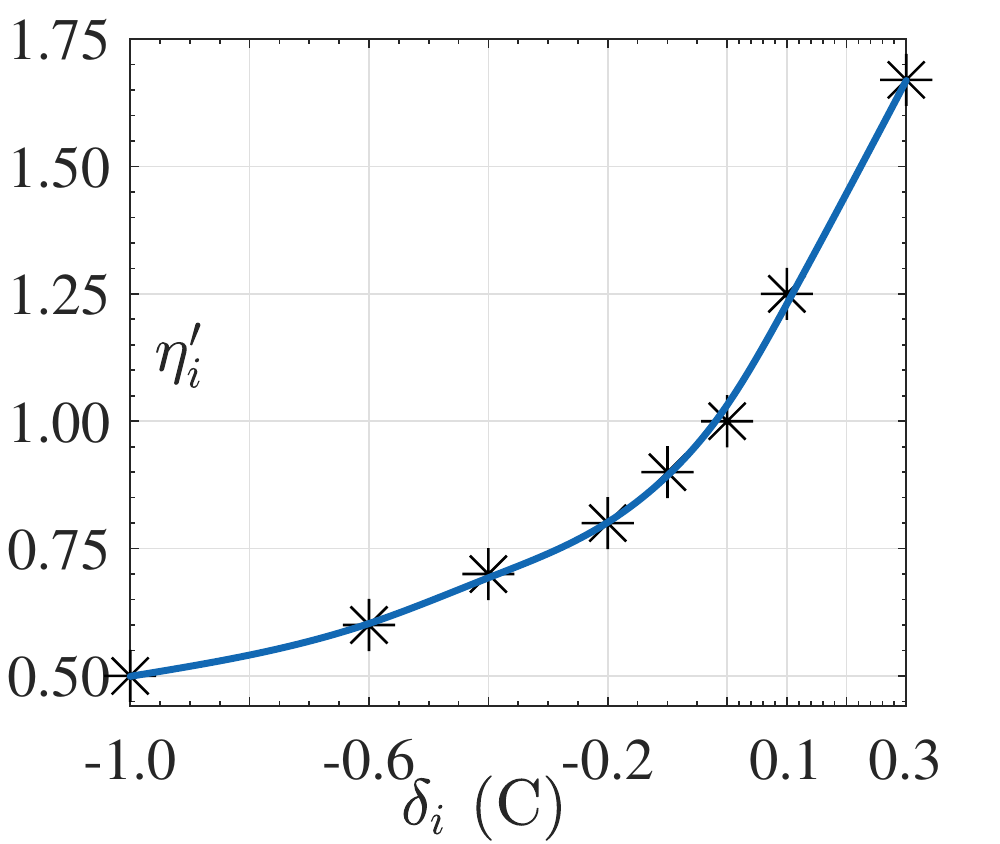}
				\label{nihe2}} 
			\caption{Example of $\eta_i$ and $\eta^{'}_i$}
		\end{figure}

Firstly, we consider the energy conversion efficiency of battery charging and discharging \cite{Cui2015Optimal}, denoted as $\eta_i(\delta_i)$ or $\eta_i$ for short, where $\eta_i\in(0,1)$. The relationship between $Q_i^{bat}$ and $\Delta_i$ is given by $Q_i^{bat}=g(\Delta_i)$, where $g(\Delta_i)$ is defined as
\begin{equation}
g(\Delta_i)= \left\{ 
       {\begin{array}{*{30}{c}}
	   \frac{1}{\eta_i}\Delta_i \, ,& \Delta_i\geq0\\
	   {\eta_i}{\Delta_i} \, ,& \Delta_i<0
	   \end{array}} \right. \, .
\end{equation}
That is, if we want to charge batteries with $\Delta_i$ kW.h, we need to actually supply $\frac{1}{\eta_i}\Delta_i$ kW.h. On the contrary, when batteries discharge $\Delta_i$ kW.h, we can actually only obtain ${\eta_i}{\Delta_i}$ kW.h. Fig. \ref{nihe1} shows a real example of $\eta_i$ \cite{StrorageChargedata}. Given that $\eta_i$ is not monotonic, we redefine $\eta_i$ as $\eta_i^{'}$ in (\ref{simply}) for simplicity.
	\begin{equation}
\eta_i^{'}= \left\{ 
{\begin{array}{*{30}{c}}
	\frac{1}{\eta_i} \, ,& \Delta_i\geq0\\
	{\eta_i} \, ,& \Delta_i<0
	\end{array}} \right.
\label{simply}
\end{equation}
		Then we have 
		\begin{equation}
		Q_i^{cons}=g(\Delta_i)=\eta_i^{'}\Delta_i
		\label{eq8}
		\end{equation}
   The graph of $\eta_i^{'}$ is shown in Fig.\ref{nihe2}, which is monotonic. It is observed that $\eta_i^{'}$ may be well fitted with a power function or an exponential function, which will be further analyzed in our simulations. 

		Secondly, another issue that we should be concerned about is that present decisions on charging or discharging will influence the future cost\cite{Cui2015Optimal}, which is denoted as potential cost. For instance, the more the power that the battery discharges at present, the more the power it will have to charge in the future, which will increase the future power cost. Accordingly, the potential cost can be defined as $-\hat{\varepsilon}_i\Delta_i$, where $\hat{\varepsilon}_i>0 $ is an evaluated parameter that can be commonly calculated as $\hat{\varepsilon}=\sum_{h=1}^{H} w_i(h)\bar{v}^s_i(h)$, where $h$ represents the indic of a future hour, $H$ represents the number of future hours, $\bar{v}^s_i(h)$ represents the predicted unit cost of power storage at the $h$th hour and $w_i(h)$ is a weight coefficient with the constraint of $\sum_{h=1}^{H} w_i(h)=1$. In addition, according to \cite{Sun2017Energy}, the influence of the current $\Delta_i$ on the future $\Delta_i(h)$ will decline with the increase of $h$. In turn, the influence of $\Delta_i(h)$ on the current $\Delta_i$ will also decline with the increase of $h$, i.e., $w_i(h_1)>w_i(h_2)$, for $\forall h_1<h_2$. Furthermore, it was shown in \cite{Sun2017Energy} that after the point $h=6$ further changes are negligible. Thus we can reasonably make $h\in\{1,2,\cdots,H\}$, where $H=6$. 		
		
		Finally, the power cost model considering storage scheduling for the $i$th data center can be extended from (\ref{q-n}) to 
		\begin{equation}
			\begin{aligned}\label{F^PC}
				&{F}^{PC}_i(\mathbf{q}_{i},\Delta_i)=\sum\limits_{n=1}^{N_i}\{ 	a_{i,n}q_{i,n}^2+p_{i,n}q_{i,n}\}-\hat{\varepsilon}_i \Delta_i\\
				&\begin{array}{r@{\quad}r@{}l@{\quad}l}
					{\rm{s.}}\  {\rm{t}}{\rm{.}}\quad & \max\{-c_i,\Delta_i^{lb}\}\leq \Delta_i \leq \min\{C_i-c_i,\Delta_i^{ub} \}\\
					&\sum\limits_{n=1}^{N_i}q_{i,n}=Q_i^{cons}+\eta_i^{'} \Delta_i\\
					& q_{i,n} \geq 0, \forall n \in \mathcal{N}_i\, .\\
				\end{array}
			\end{aligned}
		\end{equation} 
    	The first condition in (\ref{F^PC}) can be decoupled into $\Delta_i^{lb}\leq \Delta_i \leq \Delta_i^{ub}$ and $-c_i\leq \Delta_i \leq C_i-c_i$, where $\Delta_i^{lb}$ and $\Delta_i^{ub}$ are denoted as the lower bound and upper bound of $\Delta_i$, respectively. Due to the fact that an excessively high speed of charge and discharge will cause severe damage to storage devices as well as exorbitant wastes of energy \cite{StrorageCharge2}, we assume that $\Delta_i^{lb}\geq \tau^t\times(-C_i)$ and $\Delta_i^{ub}\leq \tau^t\times(0.3C_i)$. In addition, we have $-c_i\leq \Delta_i \leq C_i-c_i$, because batteries can discharge no more than the stored power and can charge no more than the remaining capacity.

\subsection{Joint Optimization of Power Cost and Workload Scheduling}\normalsize

		In this subsection, we will describe the workload scheduling model considering delay constraints, and then formulate the eco-friendly power cost minimization model with workload scheduling. We assume that the arrival rate of requests to the $i$th data center approximately obeys Poisson distribution. According to the M/M/n queuing model \cite{Rao2012Coordinated}, the average delay of tasks in the $i$th data center is given by
		\begin{equation}
			D^q_i=\frac{1}{m_i^{ac}\bar{u}_i-\lambda_i^{DC}}+\frac{1}{\bar{u}_i} \, , 
		\end{equation} 
		where $\bar{u}_i$ and $\lambda_i^{DC}$ are the average service rate per server and the average arrival rate of requests to the $i$th data center, respectively. 
		
		A lower queuing delay usually means that more servers are active, which will cause a higher electricity bill. Thus in order to study the trade-off between the power cost and the request delay, the cost function $\Phi_i$ of the $i$th data center can be defined as
		\begin{equation}
			\label{weighted function}
			\begin{aligned}
				&\Phi_i=\theta_{1,i}D^q_i+\theta_{2,i}F^{PC}_i\\ 
				=&\theta_{1,i}\left\{\frac{1}{m_i^{ac}\bar{u}_i-\lambda_i^{DC}}+\frac{1}{\bar{u}_i}\right\}\\
				+&\theta_{2,i}\left\{\sum\limits_{n=1}^{N_i}\Big{\{}a_{i,n}q_{i,n}^2+p_{i,n}q_{i,n}\Big{\}}-\hat{\varepsilon}_i \Delta_i \right\} \, ,
			\end{aligned}
		\end{equation}
		where $\theta_{1,i}>0$ and $\theta_{2,i}>0$ are weight parameters. 
		
		In addition, in order to meet the Service-Level Agreement (SLA) of the users, the transmission delay $D^t_{ji}$ between the $j$th portal server and the $i$th data center should also be considered. Since the transmission delay usually depends on the routing of the request, $\lambda_{ji}$, we define $D^t_{ji}$ as
		\begin{equation}
			D^t_{ji}=f_{ji}(\lambda_{ji}),\quad \forall j \in \mathcal{J},\forall i \in \mathcal{I}\, ,
		\end{equation}
	
		Here, $f_{ji}(\cdot)$ is an implicit function with respect to $\lambda_{ji}$, and $f(0)=0$. 
		For simplification, we denote $D^t_{i}(\boldsymbol{\lambda}_{i}^{DC})$, or $D^t_{i}$ for short, as the maximum of $D^t_{1i},\cdots,D^t_{Ji}$ according to \cite{Shao2014Optimal}, where $\boldsymbol{\lambda}_{i}^{DC}=\{\lambda_{1i},\cdots,\lambda_{Ji}\}$. Then we have 
		\begin{equation}
			\label{delay constraints}
			\frac{1}{m_i^{ac}\bar{u}_i-\lambda_i^{DC}} +\frac{1}{\bar{u}_i}+D^t_{i}\leq D \, ,
		\end{equation}
		where $D$ is the maximum delay that users can tolerate.
		
		Based on (\ref{eq1}), (\ref{eq8})--(\ref{weighted function}) and (\ref{delay constraints}), the integral cost function $\Phi$ of all back-end data centers is given by 
			\begin{subequations}\label{Cost1}
				\begin{align}
				&\mathop{\min}\limits_{\{\lambda_i^{DC},m_i^{ac},\Delta_i,\mathbf{q}_i\},\forall i\in\mathcal{I}}   \Phi=\sum\limits_{i=1}^{I}\Phi_i  \nonumber\\ 
				&\quad \quad\quad\quad\quad    =\sum\limits_{i=1}^{I}\bigg{\{}\theta_{1,i}\Big{\{}\frac{1}{m_i^{ac}\bar{u}_i-\lambda_i^{DC}}+\frac{1}{\bar{u}_i}\Big{\}} \nonumber\\
				&\quad \quad\quad\quad\quad    +\theta_{2,i}\Big{\{}\sum\limits_{n=1}^{N_i}\{a_{i,n}q_{i,n}^2+p_{i,n}q_{i,n}\}-\hat{\varepsilon}_i\Delta_i\Big{\}} \bigg{\}}\\
				&\quad \quad {\rm{s.}}\  {\rm{t}}{\rm{.}}\quad\quad      \sum\limits_{i=1}^{I}\lambda_i^{DC}=L \label{Cost1_c1}\\
				&\quad \quad \quad \quad \quad \quad                      m_{i}^{ac}\bar{u}_i-\lambda_i^{DC}\geq \frac{1}{D-\frac{1}{\bar{u_i}}-D_i^{t}} \label{couple} \\
				&\quad \quad \quad \quad \quad \quad                      \sum\limits_{n=1}^{N_i}q_{i,n}=Q_i^{cons}+\eta_i^{'} \Delta_i \label{Cost1.1} \\
				&\quad \quad \quad \quad \quad \quad                      Q_i^{cons}=\tau^t\times(m_i^{ac}s_i^{\alpha}+\beta_i) \label{Cost1_c4} \\
				&\quad \quad \quad \quad \quad \quad                      max\{-c_i,\Delta_i^{lb}\}\leq \Delta_i \leq min\{C_i-c_i,\Delta_i^{ub} \} \label{Cost1_c5} \\
				&\quad \quad \quad \quad \quad \quad                      m_i^{ac}\in\{1,2,...,M_i\},\ \lambda_i^{DC}\geq 0 \label{Cost1_c6} \\
				&\quad \quad \quad \quad \quad \quad                      q_{i,n}\geq 0,\ \forall i \in \mathcal{I}, \, \forall n \in \mathcal{N}_i \ . \label{Cost1_c7} 
				\end{align}
			\end{subequations}

 We aim to minimize $\Phi$, in order to optimize the eco-friendly use of power and delay-satisfied dispatch of workload.
	
In addition, the term $D^t_i$ in (\ref{couple}) should be further minimized as is done in (\ref{Cost2}). In fact, the definition domain of $\Phi$ defined by (\ref{Cost1_c1})--(\ref{Cost1_c7}) will be expanded with the decline of $D^t_i$, which is likely to make the minimum of $\Phi$ smaller.  
		\begin{subequations}\label{Cost2}
			\begin{align}
				&\left\{ \begin{array}{l}
					\mathop {\min }\limits_{\boldsymbol{\lambda}_{1}^{DC}} \, D_1^t(\boldsymbol{\lambda}_{1}^{DC})\\
					\cdots \\
					\mathop {\min }\limits_{\boldsymbol{\lambda}_{I}^{DC}} \, D_I^t(\boldsymbol{\lambda}_{I}^{DC})\\
				\end{array} \right.\\
				&{\rm{s.}} \ {\rm{t.}}\,\quad \sum\limits_{i = 1}^I {\lambda _{ji}}  = \lambda_j^{PC} \label{Cost2_c1}\\
				&\quad \ \ \quad \sum\limits_{j = 1}^J {\lambda _{ji}}  = \lambda _i^{DC} \, , \forall j \in \mathcal{J}, \, \forall i \in \mathcal{I} \, . \label{Cost2_c2}
			\end{align}
		\end{subequations}
	
     Finally, the whole eco-friendly power cost minimization model with workload scheduling for geo-distributed data centers can be defined as
     
     \begin{equation}
     \label{P&W}
     \begin{aligned}
   & \qquad \left\{ \begin{array}{l}
     \mathop{\min}\limits_{\{\lambda_i^{DC},m_i^{ac},\Delta_i,\mathbf{q}_i\},\forall i\in\mathcal{I}}   \Phi \\
     \qquad \mathop {\min }\limits_{\boldsymbol{\lambda}_{1}^{DC}} \, D_1^t(\boldsymbol{\lambda}_{1}^{DC})\\
     \qquad \textbf{$\cdots$} \\
    \qquad \mathop {\min }\limits_{\boldsymbol{\lambda}_{I}^{DC}} \, D_I^t(\boldsymbol{\lambda}_{I}^{DC})  
     \end{array} \right.\\
    &{\rm{s.}} \ {\rm{t.}}\, \quad ({\rm{\ref{Cost1_c1})\, to \, (\ref{Cost1_c7}), \, (\ref{Cost2_c1}) \, and \, (\ref{Cost2_c2}})}\, . \\
     \end{aligned}
     \end{equation}	
  	For convenience, we denote Problem (\ref{P&W}) as Problem P\&W, which is a multi-objective programming problem, and denote Problem (\ref{Cost1}) and Problem (\ref{Cost2}) as Problem P and Problem W, respectively.

\section{Solution and Analysis} 

 As Problem P\&W is a multi-objective programming problem with integer constraints, it is difficult to solve. But if we assume that $D^t_i$ in (\ref{Cost1}c) is a known constant, then Problem P\&W can be simplified into the easier Problem P. Thus in this section, we first regard $D^t_i$ as a known constant to simplify Problem P\&W as Problem P. We propose a Sequential Convex Programming (SCP) algorithm, shown in Algorithm \ref{Algorithm2}, to find the non-integer solutions of Problem P. Then we propose a low-complexity searching method, shown in Algorithm \ref{Algorithm3}, to seek for the quasi-optimal mixed-integer solutions of Problem P. This approach can obtain a quasi-optimal solution, which is very close to the optimal solution obtained by the Branch and Bound (B\&B) method, while converging much faster. Finally, we reconsider $D^t_i$ as a variable and give the equivalence condition of Problem P and Problem P\&W.

\subsection{Non-integer Solution of simplified Problem P$\&$W}

In this subsection, we just regard $D^t_i$ in (\ref{Cost1}c) as a known constant and propose an SCP algorithm to find the non-integer solution of Problem P, the simplified version of Problem P\&W. We denote the continuity-relaxed version of Problem P as Problem P$^*$. It is observed that Problem P$^*$ is non-convex because of the non-linear equality constraint (\ref{Cost1}d) \cite{boyd2004convex}, in which $\eta_i^{'}$ is a non-linear function of $\Delta_i$ rather than a constant. As is seen, (\ref{Cost1}d) is mainly related to the second part of $\Phi$, which is defined in (\ref{q-n}). Therefore, we will first solve (\ref{q-n}) and then plug the solution of $F_i^{PC}(\cdot)$ into (\ref{Cost1}) to eliminate variables $q_{i,n},\forall i \in \mathcal{I}, \forall n \in \mathcal {N}_i$. By this, Problem P$^*$ can be transformed to sequential convex problems and its globally optimal solution can be obtained by our proposed SCP algorithm.

First, we ignore the constraint $q_{i,n}\geq 0$ and establish the Lagrange Dual Function of (\ref{q-n}) for the $i$th data center as   
\begin{equation}\label{Lagrange}
	L_i(\mathbf{q}_i,\mu_i)=\sum\limits_{n=1}^{N_i}\{a_{i,n}q_{i,n}^2+p_{i,n}q_{i,n}\}-
	\mu_i^L\Big{\{}\sum\limits_{n=1}^{N_i}q_{i,n}-Q_i^{cons}-Q_i^{bat}\Big{\}} \, ,
\end{equation}
where $\mu_i^{L}$ is the Lagrange factor. According to the Lagrange Multiplier method \cite{boyd2004convex}, we set $\frac{\partial L_i}{\partial q_{i,n}}=0$ and $\frac{\partial L_i}{\partial \mu_i}=0$ for $\forall n \in \mathcal{N}_i$. Then we can obtain the Lagrange Dual Solution of (\ref{q-n}) without $q_{i,n}\geq 0$ as
\begin{equation}\label{q-n solution}
	\begin{aligned}
		q_{i,n}^{*L}&=\frac{v_i^{*L}-p_{i,n}}{2a_{i,n}}\\
		&=\frac{2(Q_i^{cons}+Q_i^{bat})+Y_i-p_{i,n}X_i}{2a_{i,n}X_i} \, .
	\end{aligned}
\end{equation}
The corresponding optimal function value of $F^{EC}_i(\cdot)$ in (\ref{q-n}) can be given by 
\begin{equation}\label{simplify2}
\begin{aligned}
F^{PC}_i(\mathbf{q}_i^{*L})=\frac{(Q_i^{cons}+Q_i^{bat})^2+Y_i(Q_i^{cons}+Q_i^{bat})+\frac{1}{4}(Y_i^2-X_iZ_i)}{X_i} \, ,\\
\end{aligned}
\end{equation} 
where
\begin{equation}\label{XY}
	\begin{aligned}
		X_i\, &=\sum\limits_{n=1}^{N_i}\frac{1}{a_{i,n}}\, ,\quad	Y_i\, =\sum\limits_{n=1}^{N_i}\frac{p_{i,n}}{a_{i,n}}\, ,\quad
		Z_i\,=\sum\nolimits_{n=1}^{N_i}\frac{(p_{i,n})^2}{a_{i,n}}\, ,\\
		v_i^{*L}&=\mu_i^{L*}=\frac{2(Q_i^{cons}+Q_i^{bat})+Y_i}{X_i}\, . \\
	\end{aligned}
\end{equation} 
Here, $v_i^{*L}$ is the marginal cost of power for the $i$th data center when $q_{i,n}=q_{i,n}^{*L}$ for $\forall n \in \mathcal{N}_i$, which is independent of $n$. Note that $X_i>0$, $Y_i>0$, $Z_i>0$ and $v_i^{*L}>0$ since $a_{i,n}>0$, $p_{i,n}>0$ and $Q_i^{cons}+Q_i^{bat}\geq 0$. 

\begin{figure}[htbp]
	\centering
	\includegraphics[width=5.3cm]{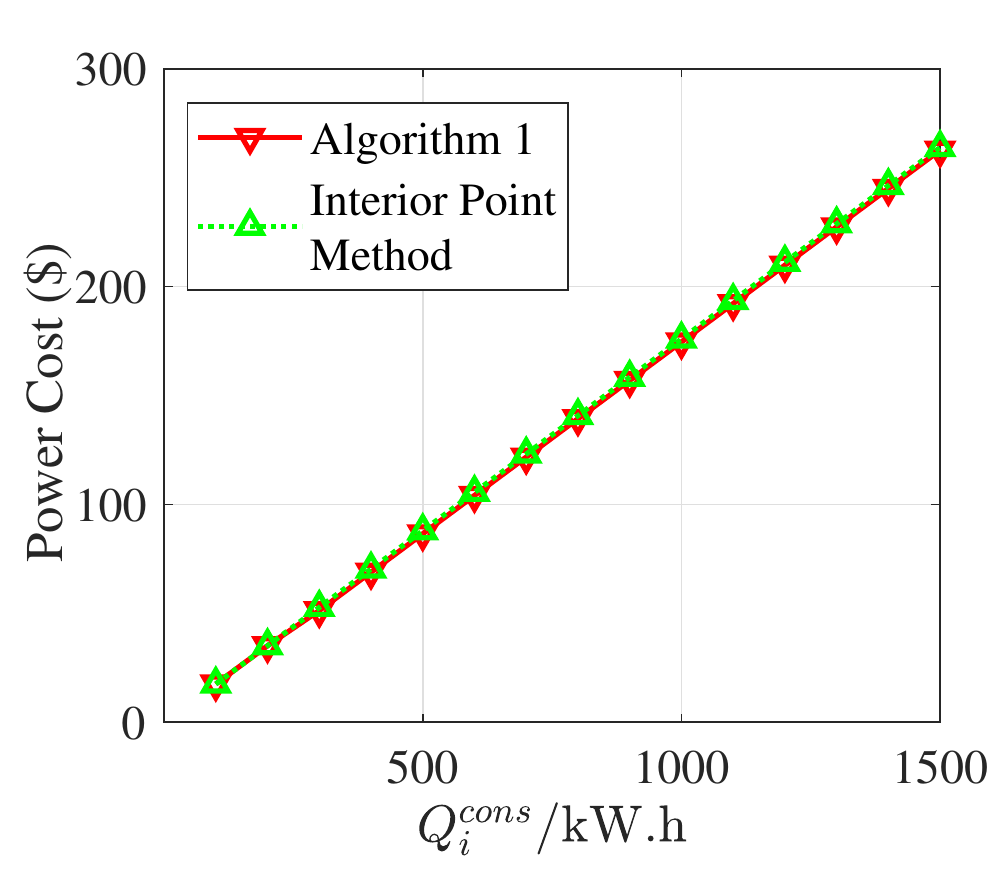}
	\caption{Comparison of Algorithm 1 and Interior Point Method}
	\label{erro1}
\end{figure}

Secondly, we reconsider the inequality constraint $q_{i,n}\geq 0$. It is proved in Appendix A that for the $q_{i,n}$ whose optimal Lagrange solution $q_{i,n}^{*L}$ obtained by Eq. (\ref{q-n solution}) is less than zero, its optimal solution with the constraint $q_{i,n}\geq0$ in problem (\ref{q-n}) is zero. Based on this, we propose Algorithm \ref{Algorithm1} to solve the optimal $\mathbf{q}_i$ in (\ref{q-n}) for the $i$th data center. We denote $\mathcal{N}_i^{-}$ as a subset of $\mathcal{N}_i$, where $q_{i,n}^{*L}<0$ for $\forall n \in \mathcal{N}_i^{-}$, and $\mathcal{N}_i=\mathcal{N}_i^{-}\cup\widetilde{\mathcal{N}}_i^{-}$. In Algorithm \ref{Algorithm1}, we make $\widetilde{\mathcal{N}}_i^{-}=\mathcal{N}_i$ at first, and then repeat the following two steps until all $q_{i,n}^{*L},n\in\widetilde{\mathcal{N}}_i^{-}$ are no less than zero: $(\romannumeral1)$ moving the indic $n$ of $q_{i,n}^{*L}<0$ from the subset $\widetilde{\mathcal{N}}_i^{-}$ to $\mathcal{N}_i^{-}$, and $(\romannumeral2)$ recalculating the remaining $q_{i,n}$ according to (\ref{q-n solution}) and (\ref{XY}). Finally, we make $q_{i,n}^{*}=0,\forall n \in \mathcal{N}_i^{-}$ and $q_{i,n}^{*}=q_{i,n}^{L*},\forall n \in \widetilde{\mathcal{N}}_i^{-}$. It is proved in Appendix A that Algorithm \ref{Algorithm1} can obtain the optimal solution of Problem (\ref{q-n}). In addition, Fig. \ref{erro1} shows that the optimal $F^{PC}_i(\cdot)$ solved by Algorithm \ref{Algorithm1} and by the Interior Point method are almost identical. According to \cite{Sun2017Energy}, Algorithm \ref{Algorithm1} converges faster than the Interior Point method when solving $\mathbf{q}_i$.
\begin{algorithm}\small
	\caption{Obtain Optimal $\mathbf{q}_i$ in (\ref{q-n})}
	\label{Algorithm1}  	 
	\begin{algorithmic}[1]
		\State  \textbf{Initialize} $N_i$, $\mathcal{N}_i$, $\gamma_{i,n}$, $p_{i,n}$, $Q_i^{cons}$ and $Q_i^{bat}$ for the $i$th data center. Make ${\mathcal{N}}_i^{-}=\emptyset$ and $\widetilde{\mathcal{N}}_i^{-}=\mathcal{N}_i$.
		\State  $~~$\textbf{Repeat} on the renewed subset $\widetilde{\mathcal{N}}_i^{-}$  
		\State  $~~~~$Build the Lagrange Dual Function as (\ref{Lagrange})
		\State  $~~~~$Calculate $X_i$, $Y_i$ and $Z_i$ according to (\ref{XY})
		\State  $~~~~$Calculate $q_{i,n}^{*L}$ as (\ref{q-n solution})
		\State  $~~~~$\textbf{If} there is any $q_{i,n}^{*L}<0,n\in \widetilde{\mathcal{N}}_i^{-}$ \textbf{then}
		\State  $~~~~~~~$Move all indics $n$ of $q_{i,n}^{*L}<0$ from $\widetilde{\mathcal{N}}_i^{-}$ to $\mathcal{N}_i^{-}$
		\State  $~~~~$\textbf{End If} 
		\State  $~~$\textbf{Until} No new indic $n$ is moved into $\mathcal{N}_i^{-}$.
		\State  Make $q_{i,n}^{*}=0$ for $\forall n \in \mathcal{N}_i^{-}$ and $q_{i,n}^{*}=q_{i,n}^{L*}$ for $\forall n \in \widetilde{\mathcal{N}}_i^{-}$. 
	\end{algorithmic}
\end{algorithm} 

Thirdly, we plug the solution of $F_i^{PC}(\cdot)$, shown in (\ref{simplify2}), into (\ref{Cost1}) to eliminate the non-linear constraint. Then we have 
\begin{subequations}\label{Cost1-equal}
	\begin{align}
		\mathop{\min}\limits_{\{\lambda_i^{DC},m_i^{ac},\Delta_i,i\in\mathcal{I}\}}  
		& \quad  \Phi=\sum_{i = 1}^{I} \Phi _i \nonumber\\
		\mathop{\Phi_i}\limits_{\{\lambda_i^{DC},m_i^{ac},\Delta _i\}}=&\theta_{1,i}\Big{\{}\frac{1}{m_i^{ac}\bar{u}_i-\lambda_i^{DC}}+\frac{1}{\bar{u_i}}\Big{\}} \nonumber\\
		+&\theta_{2,i}\Big{\{}\frac{(Q_i^{cons}+\eta_i^{'}\Delta_i)^2+Y_i(Q_i^{cons}+\eta_i^{'}\Delta_i)+W_i}{X_i}-\hat{\varepsilon}_i\Delta_i\Big{\}}\\
		\quad{\rm{s.}}\  {\rm{t}}{\rm{.}}\quad 
		& \sum\limits_{i=1}^{I}\lambda_i^{DC}=L\\
		& m_{i}^{ac}\bar{u}_i-\lambda_i^{DC}\geq \frac{1}{D-\frac{1}{\bar{u_i}}-D_i^t}\\
		& Q_i^{cons}=\tau^t\times(m_i^{ac}s_i^{\alpha}+\beta_i)\\
		& max\{-c_i,\Delta_i^{lb}\}\leq \Delta_i \leq min\{C_i-c_i,\Delta_i^{ub} \}\\
		& m_i^{ac}\in\{1,2,\cdots,M_i\}, \lambda_i^{DC}\geq 0,\ \forall i \in \mathcal{I} \, ,
	\end{align}
\end{subequations}
where $W_i=\frac{1}{4}(Y_i^2-X_iZ_i)$. We denote problem (\ref{Cost1-equal}) as Problem P1 and denote the continuity-relaxed version of problem (\ref{Cost1-equal}) as Problem P1$^*$. We find that the feasible region of  Problem P1$^*$ turns into a convex set after eliminating the nonlinear equality constraint. According to \cite{boyd2004convex}, only if the objective function (\ref{Cost1-equal}a) is also convex can  Problem P1$^*$ be proved to be a convex problem. In Appendix B, we prove that (\ref{Cost1-equal}a) is convex when the fitted function of $\eta_i^{'}$ meets the condition shown in Proposition 4, which is easily to be met according to simulations. Therefore, we just regard (\ref{Cost1-equal}a) as a convex function in this paper, so that  Problem P1$^*$ is a convex programming problem.

Based on all of the above, we propose the SCP algorithm to solve  Problem P$^*$. The details of the SCP algorithm are shown in Algorithm \ref{Algorithm2}, which is proved to obtain the optimal non-integer solution of Problem P$^*$ in Appendix A. In the SCP algorithm, every time we solve a new version of $\{q_{i,n}\}$, we plug $F_i^{PC}(\cdot)$ into problem (\ref{Cost1-equal}) and obtain a new Problem P1$^*$, which can be solved by standard convex programming methods, such as Sequential Quadratic Programming (SQP). Algorithm \ref{Algorithm1} and Algorithm \ref{Algorithm2} adopt similar architectures and the main differences between them include $(\romannumeral1)$ Lines 3--4 in Algorithm \ref{Algorithm2} to obtain and solve a new version of Problem P1$^*$, and $(\romannumeral2)$ Lines 10--12 in Algorithm \ref{Algorithm2} analyzed in Appendix A. We conclude that our proposed SCP algorithm can obtain the global optimal solution of Problem P$^*$ by solving several instances of convex Problem P1$^*$.
\begin{algorithm}\small	
	\caption{Sequential Convex Programming Algorithm}
	\label{Algorithm2}   
	\begin{algorithmic}[1]
		\State  \textbf{Initialize} relevant parameters. Make ${\mathcal{N}}_i^{-}=\emptyset$ and $\widetilde{\mathcal{N}}_i^{-}=\mathcal{N}_i$ for $\forall i \in \mathcal{I}$.
		\State  $~~$\textbf{Repeat}  
		\State  $~~~~$Calculate $X_i$, $Y_i$ and $Z_i$ on subset $\widetilde{\mathcal{N}}_i^{-}$ according to (\ref{XY}) $~~~~~$ for $\forall i \in \mathcal{I}$, and plug them into Problem P1$^*$.
		\State  $~~~~$Solve Problem P1$^*$ with convex programming method, $~~~~~$ such as SQP.
		\State  $~~~~$Calculate $q_{i,n}^{*L}$ and $\nu_i^{*L}$ as in (\ref{q-n solution}) and (\ref{XY}), respectively, $~~~~~$ on subset $\widetilde{\mathcal{N}}_i^{-}$ for $\forall i \in \mathcal{I}$.
		\State  $~~~~$\textbf{For} $i=1:I$
		\State  $~~~~~~$\textbf{If} there is any $q_{i,n}^{*L}<0,n\in \widetilde{\mathcal{N}}_i^{-}$ \textbf{then}
		\State  $~~~~~~~~$Move all indices $n$ of $q_{i,n}^{*L}<0$ from $\widetilde{\mathcal{N}}_i^{-}$ to $\mathcal{N}_i^{-}$
		\State  $~~~~~~$\textbf{End If}
		\State  $~~~~~~$\textbf{If} there is any $p_{i,n}<\nu_i^{*L},n\in \mathcal{N}_i^{-}$ \textbf{then}
		\State  $~~~~~~~~$Move all indices $n$ of $p_{i,n}<\nu_i^{*L}$ back to $\widetilde{\mathcal{N}}_i^{-}$
		\State  $~~~~~~$\textbf{End If}
		\State  $~~~~$\textbf{End For} 
		\State  $~~$\textbf{Until} No $n$ is newly moved into $\mathcal{N}_i^{-}$ or $\widetilde{\mathcal{N}}_i^{-}$ for $\forall i\in\mathcal{I}$
		\State  Output the renewed $\lambda_i^{DC}$, $m_i^{ac}$, $\Delta_i$, $q_{i,n\in \mathcal{N}_i^{-}}^{*}=0$ and $q_{i,n\in \widetilde{\mathcal{N}}_i^{-}}^{*}=q_{i,n}^{L*}$ for $\forall i\in\mathcal{I}$.	
	\end{algorithmic}
\end{algorithm}  
\subsection{Mixed-integer Solution of Problem P$\&$W} 
In this subsection, we consider the integer constraints (\ref{Cost1-equal}f) and propose Algorithm \ref{Algorithm3} to find a quasi-optimal mixed-integer solution of Problem P at first.

 Considering that B\&B method is the most common integer programming algorithm and is the foundation of others \cite{leyffer2001integrating} \cite{Westerlund1995An} \cite{Wei2015Outer}, we firstly combine B\&B with our proposed SCP algorithm, denoted as BB-SCP, to seek for the optimal mixed-integer solution of Problem P. In BB-SCP, we need to solve Problem P$^*$ by the SCP algorithm in each branch (iteration). Simulation results show that the times of branching for the B\&B algorithm will increase rapidly with the growth of $I$, the number of integer-limited variables $m_i^{ac}$. For example, when $I$ reaches about 8, the number of branches will grow to more than 5000. Although many studies tried to improve the convergence speed of B\&B algorithm \cite{leyffer2001integrating} \cite{quesada1992lp}, it is still a severe problem. Therefore, we propose Algorithm \ref{Algorithm3} to replace BB-SCP and seek for a quasi-optimal mixed-integer solution of Problem P .

Algorithm \ref{Algorithm3} mainly refers to the rounding strategy towards integer-limited $m_i^{as}$. In detail, we first solve Problem P$^*$ by our proposed SCP algorithm and make each $m_i^{ac}$ equal to the closest integer. Then we adjust the obtained integer-values of $m_i^{as}$ and recalculate the remaining variables by the SCP algorithm again. Table \ref{table1} shows the quasi-optimal solutions obtained by Algorithm \ref{Algorithm3} and the gaps between the quasi-optimal solutions and the optimal solutions obtained by the BB-SCP method, where the gaps of $\Phi$ can be seen to be very small. According to our simulation, we can conclude that Algorithm \ref{Algorithm3} can not only obtain a quasi-optimal mixed-integer solution of Problem P with a tiny loss, but also needs to call the SCP algorithm only twice instead of thousands of times as required by the BB-SCP algorithm. In addition, we also find from Table \ref{table1} that the gaps of total power cost are negligible, and the average $D_i^{q}$ remains small despite the non-ignorable gaps.

\begin{algorithm}\small	
	\caption{Quasi-optimal Mixed-integer Solution}
	\label{Algorithm3}   
	\begin{algorithmic}
		\State  Initialize relevant parameters, solve Problem P$^*$ with our proposed SCP algorithm, and then adjust the values of $m_i^{ac}$ as follows.
		\State  $~$ $m_i^{Int}=$round$(m_i^{ac})$ 
		\State  $~$ $gap_1=\sum\limits_{i\in\mathcal{I}}\left\{m_i^{Int}\bar{u}_i-m_i^{ac}\bar{u}_i\right\} $
		\State  $~$ $\mathbf{d}=\mathbf{m}^{Int}-\mathbf{m}^{ac}$, where $\mathbf{m}^{Int}=\{m_1^{Int}...,m_I^{Int}\}$ and $\mathbf{m}^{ac}=$
		\State  $~$ $\{m_1^{ac},...,m_I^{ac}\} $
		\State  $~$ Denote $d^s_k$ as the $k$th smallest element of vector $\mathbf{d}$
		\State  $~$ Denote $I^d_k$ as the indic of $DC_i$ for $d^s_k$
		\State  $~~$\textbf{If} $gap_1<\omega_1$ \textbf{then}
		\State  $~~~~$ $num_1$=ceil\{abs($\frac{\sum_{i\in\mathcal{I}}m_i^{ac} \bar{u}_i-L}{\bar{u}}$)\}, where $\bar{u}=\frac{\sum_{i\in\mathcal{I}}\bar{u}_i}{I}$ 
		\State  $~~~~$ \textbf{For} $k=1:num_1$ 
		\State  $~~~~~~$ \textbf{If} $d^s_k<\omega_2$ \textbf{then}
		\State  $~~~~~~~~$  Add 1 to $m_{I^d_k}^{Int}$
		\State  $~~~~~~$ \textbf{End If}
		\State  $~~~~$ \textbf{End For}
		\State  $~~$\textbf{End If}	
		\State  $~$ $gap_2=\sum\limits_{i\in\mathcal{I}}\left\{m_i^{Int}-m_i^{ac}\right\} $ and $num_2 =$round$(\left|gap_2\right|)$
		\State  $~~$\textbf{If} $gap_2<\omega_3$ \textbf{then}
		\State  $~~~~$ \textbf{For} $k=num_1+1:num_1+num_2$ 
		\State  $~~~~~~$Add 1 to $m_{I^d_k}^{Int}$ 
		\State  $~~~~$ \textbf{End For}
		\State  $~~$\textbf{End If}	
		\State  $~~$\textbf{If} $gap_2>\omega_4$ \textbf{then}
		\State  $~~~~$ \textbf{For} $k=1:num_2$ 
		\State  $~~~~~~$Subtract 1 from $m_{I^d_{I-k+1}}^{Int}$ 
		\State  $~~~~$ \textbf{End For}
		\State  $~~$\textbf{End If} 
		\State  $~$ Add 1 to $m_i^{Int}$ which equal to 0 
		\State  $~$ Make $\mathbf{m}^{ac}=\mathbf{m}^{Int}$. 
		\State  Fix $\mathbf{m}^{ac}$ and seek for optimal solutions of the remaining variables with the SCP algorithm.		
	\end{algorithmic}
\end{algorithm} 

\begin{table}[htb] \small
	\centering
	\caption{Gaps between optimal and quasi-optimal solutions}
	\begin{tabular}{|l|c|c|c|c|c|c|}
		\hline
       \multirow{2}{*}{$I$} &\multicolumn{2}{c|}{$\Phi$}&\multicolumn{2}{c|}{Total power cost} &\multicolumn{2}{c|}{Average ${D}^q_i$}\\
       \cline{2-7}
                            & Value & Gap              & Value & Gap                 & Value & Gap \\
       \hline
        2     & 1279.27   &  0.10         & 1350.36   &  0.22          & 0.100   & -0.069\\
        4     & 2491.27   &  0.19         & 2596.17   &  0.44          & 0.100   & -0.067\\
        6     & 3392.82   &  0.01         & 3495.04   &  0.10          & 0.150   & -0.014\\
        8     & 4592.81   &  0.05         & 4657.17   & -0.19          & 0.200   &  0.035\\
        10    & 5987.30   &  0.00         & 5782.49   &  0.02          & 0.167   & -0.001\\       
        12    & 7056.10   &  0.02         & 7063.86   &  0.17          & 0.150   & -0.016\\
        14    & 8128.23   &  0.01         & 8157.40   & -0.11          & 0.175   & -0.009\\
        16    & 9041.70   &  0.00         & 9104.80   &  0.06          & 0.160   & -0.004\\
		\hline
	\end{tabular}
	\label{table1}  
\end{table}

Finally, we regard $D_i^t$ as a variable instead of a constant to solve the original version of Problem P\&W shown in (\ref{P&W}). It can be seen from Table \ref{table1} that all $D_i^q$ solved in Problem P are always less than 0.3s, while the maximum tolerable delay $D$ is set to 5 s. That is, benefiting from making $D_i^q$ a part of the objective function (\ref{Cost1}a), $D_i^q$ is always optimized to be as small as possible rather than to be close to the upper bound as in \cite{Shao2014Optimal}--\cite{Yao2014Power}. Therefore, we assume that the transmission delay $D_i^t$ will never exceed $D-D_i^{q,up}$ for $\forall i \in \mathcal{I}$ and set $D_i^{q,up}$ to 1 s in this paper, where $D_i^{q,up}$ is the maximum $D_i^q$. Then (\ref{Cost1}c) becomes a slack constraint and Problem P\&W is equal to its simplified version, Problem P, and can be solved by our proposed algorithms.

\section{Numerical Analysis}
To verify our model, we will use Matlab to solve Problem P by our proposed Algorithm \ref{Algorithm3} and give the performance analysis in terms of four aspects, including $(1)$ the impact of PIF, $(2)$ the behavior of storage scheduling, $(3)$ the analysis of workload scheduling, and $(4)$ the reduction of power cost and the improvement of clean power usage. In our model, we select the default parameters according to Table \ref{parameter}. Specifically, we assume that each data center can buy $N_i=3$ kinds of power, including the thermal power (TP) supplied by bulk power grid, and the wind power (WP) and solar power (SP) supplied by renewable power suppliers. The corresponding electricity prices\footnote{All electricity prices used in this paper are set according to those in China, and converted from $\yen$ to \$ with exchange rate 0.148.} always change at different hours and in different areas. 
\begin{table}[!htb] \small
	\centering
	\caption{Parameters Settings}
	\begin{tabular}{lccc}
		\hline
		\cline{1-2} 
		\textbf{Parameters}                               &     \textbf{Values} \\
		Upper bound of power consumption, $P_i^{max}$           &      1 MW \\
		Capacity of storage, $C_i$                        &      40\%$P_i^{max}$$\,$--$\,$60\%$P_i^{max}$ \\
		Power of active server, $s_i^{\alpha}$  	      &      0.4 kW$\,$--$\,$0.7 kW \\
		Basic power consumption, $\beta_i$                &      40 kW$\,$--$\,$60 kW  \\
		Processing speed of active server, $\bar{u}_i$    &      80 requests per second \\
		Upper bound of delay, $D$                         &      2s \\
		Upper bound of transmission delay                 &      2s \\
		Power factor, $f^P$                   &      $\frac{1}{\tau^tP_i^{max}}$ \\
		Pollution factor of TP, $\gamma_{i,1}$       &      0.5 \\
		Pollution factor of WP, $\gamma_{i,2}$       &      0.4 \\
		Pollution factor of SP, $\gamma_{i,3}$       &      0.3 \\
		Price of TP, $p_{i,1}$                            &      0.04$\,$--$\,$0.12 kW.h \\
		Price of WP, $p_{i,2}$                            &      0.08$\,$--$\,$0.14 kW.h  \\
		Price of SP, $p_{i,3}$                            &      0.10$\,$--$\,$0.18 kW.h	 \\
		\hline
	\end{tabular}
	\label{parameter}  
\end{table}

\subsection{Effect of the Pollution Index Function} 
\begin{figure}[htp]
	\centering
	\subfloat[No $f^P$]{\includegraphics[width=4.3cm]{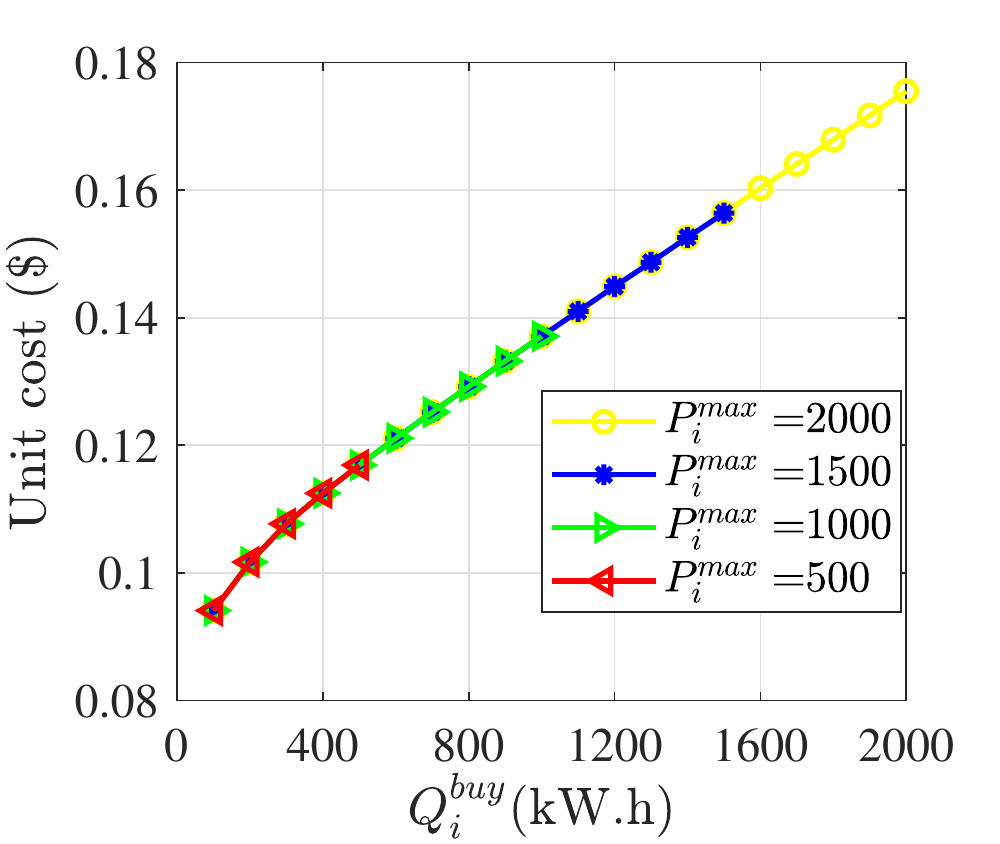}
		\label{gamma_1}} 
	\hfill
	\subfloat[$f^P=\tau^t\times P_i^{max}$]{\includegraphics[width=4.3cm]{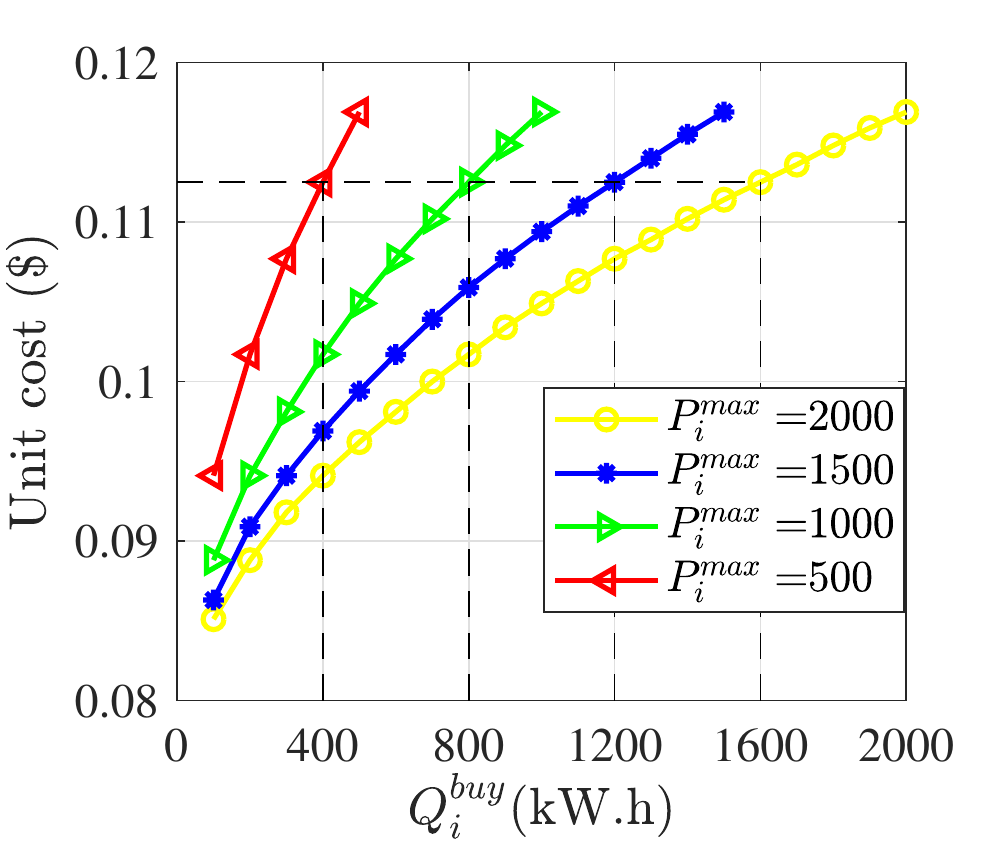}
		\label{gamma_2}} 
	\caption{The effect of power factor $f^P$ on unit cost}
\end{figure}

First, we aim to show the effect of the power factor $f^P$ in PIF, which is inversely proportion to the maximum power $P_i^{max}$ of the $i$th data center. We vary $P_i^{max}$ from 500 kW to 2000 kW and solve problem (\ref{q-n}) by Algorithm \ref{Algorithm1}. Fig. \ref{gamma_1} shows the unit costs when replacing $f^P$ with a constant 500 for comparison. We can see that the unit costs are increasing with the growth of $Q_i^{buy}$, and the data center of larger $P_i^{max}$ always pays a higher unit cost when the same fraction power of $P_i^{max}$ is bought. 
Fig. \ref{gamma_2} shows the unit costs considering $f^P=\tau^t\times P_i^{max}$, where the unit costs are the same when the ratio $\frac{Q_i^{buy}}{\tau^t\times P_i^{max}}$ is fixed and the $P_i^{max}$ ranges from 500 kW to 2000 kW. It shows that the increment of the unit cost depends on the ratio of the bought power to the maximum power after introducing the power factor $f^P$, which can relieve the excessive increases of unit costs especially for large data centers. 

Moreover, Table \ref{table_gamms} shows the solutions of $m_i^{ac}$ in Problem P when $f^P=500$ and $f^P=\tau^t\times P_i^{max}$, respectively, in which cases the electricity prices in different areas are assumed to be identical. It is observed that when replacing $f^P$ with 500, the number of active servers $m_i^{ac}$ in different data centers are similar, but the more the total servers $M_i$, the smaller the fraction of the active servers. Conversely, the fractions of active servers are nearly the same when $f^P=\tau^t\times P_i^{max}$, because the unit cost depends on the ratio of $Q_i^{buy}$ to $\tau^t\times P_i^{max}$ instead of $Q_i^{buy}$ alone. Therefore, the servers in different scale data centers can be fairly used after introducing the power factor $f^P$. 
 \begin{table}[htb] \small
 	\centering
 	\caption{The effect of Power factor $f^P$ on $m_i^{ac}$ when $I=6$ and $L=60\%L_{max}$}
 	\begin{tabular}{|l|c|c|c|c|c|}
 		\hline
 		\multirow{2}{*}{$i$} &\multirow{2}{*}{$M_i$} &\multicolumn{2}{c|}{No $f^P$}&\multicolumn{2}{c|}{With $f^P$} \\
 		\cline{3-6}
 		&	& $m_i^{ac}$ & Use Ratio              	& $m_i^{ac}$ & Use Ratio                \\
 		\hline
 		1     & 2200   & 2150         & 97.73\%   & 1320          & 60.00\%\\
 		2     & 2600   & 2182         & 83.92\%   & 1560          & 60.00\% \\     
 		3     & 3000   & 2215         & 73.83\%   & 1801          & 60.03\% \\
 		4     & 4600   & 2346         & 51.00\%   & 2759          & 59.98\% \\
 		5     & 5000   & 2378         & 47.56\%   & 3003          & 60.06\% \\		  
 		6     & 5400   & 2411         & 44.65\%   & 3239          & 59.98\% \\
 		\hline
 	\end{tabular}
 	\label{table_gamms}  
 \end{table}

\begin{figure}[htbp]
	\centering
	\subfloat[Unit cost]{\includegraphics[width=4.35cm]{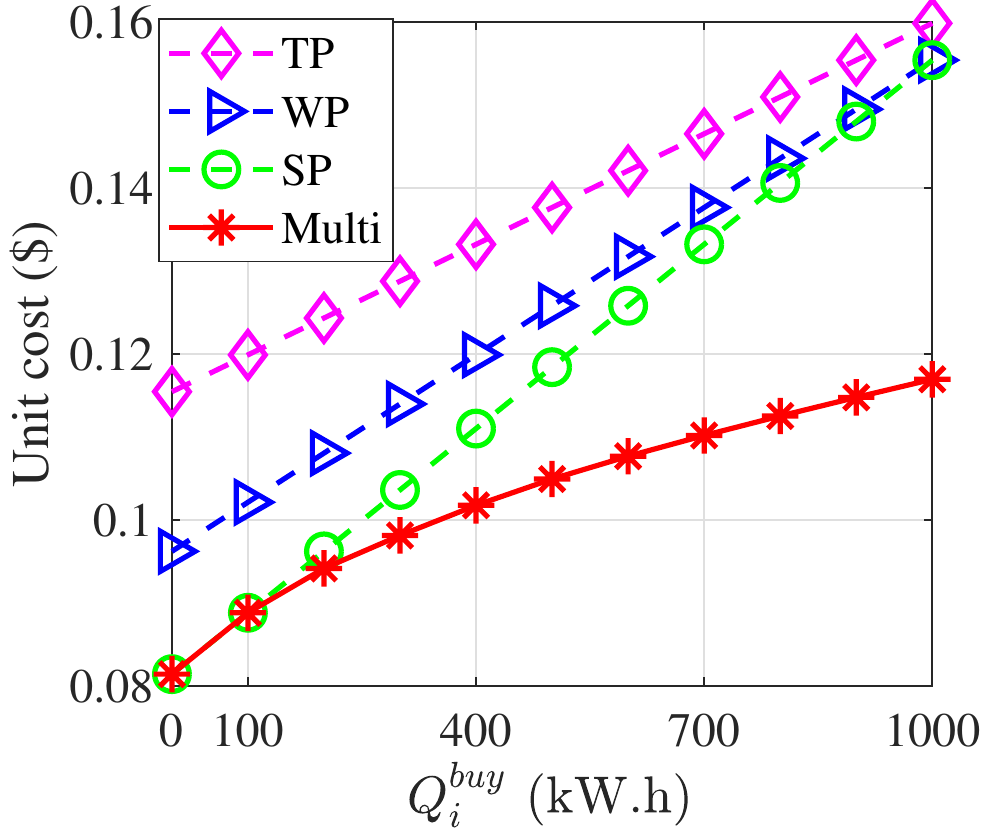}
		\label{PollutionIndex}} 
	\hfill
	\subfloat[Ratio of clean power]{\includegraphics[width=4.25cm]{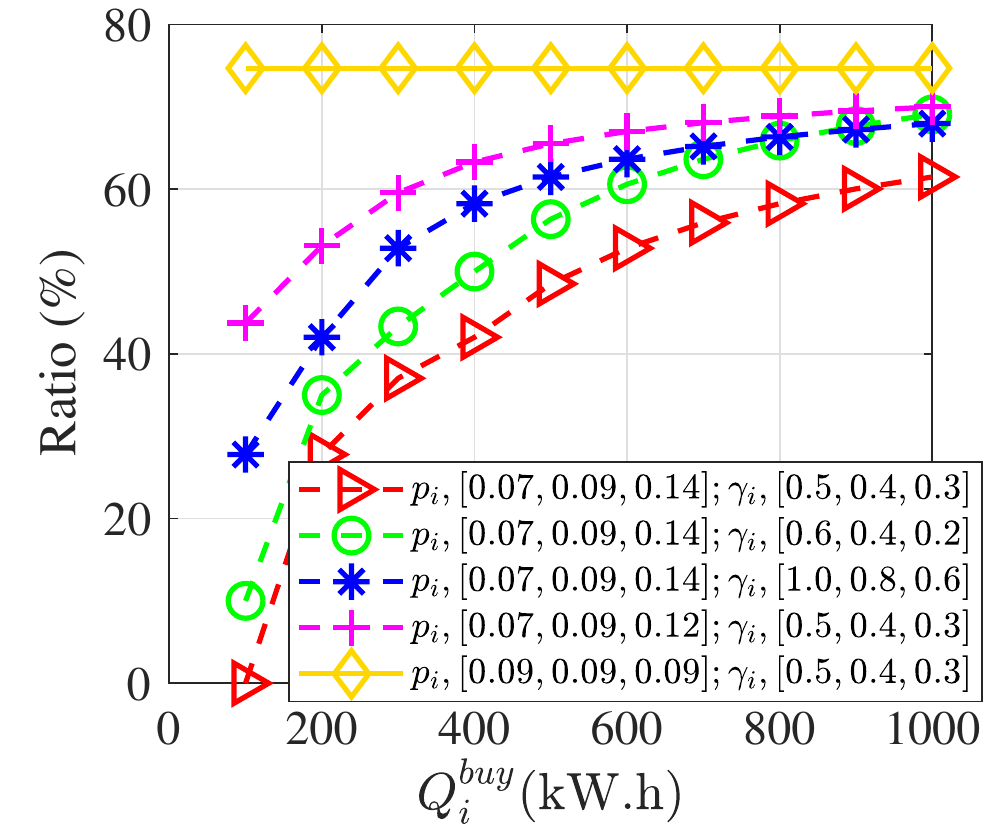}
		\label{gamma_5}} 
	\caption{Effect of PIF on power selection.}
\end{figure}

Secondly, we aim to verify the effect of PIF on the selection of power sources. Fig. \ref{PollutionIndex} shows the unit costs for a 1-MW data center when buying power from TP, WP, SP and multi-sources in one time slot, respectively. It is observed that the unit cost increases with the growth of $Q_i^{buy}$. This means that the more the power we buy, the higher the unit cost we will pay, which can lead to the increasingly fast growth of total power costs and encourage power savings in order to reduce costs. In addition, we can see that the unit costs in the cases of buying power from TP, WP or SP alone are always higher than those when buying multi-source power, which is caused by the quadratic term in PIF. Therefore, our proposed PIF can encourage users to buy power from multiple suppliers rather than single suppliers. 

Fig. \ref{gamma_5} shows the ratio of clean power, including WP and SP, to the total bought power $Q_i^{buy}$. It can be observed that the fraction of clean power will increase with the growth of $Q_i^{buy}$ when $p_{i,n}, \forall n \in \mathcal{N}_i$ are different, which means that the more the power we buy, the higher the fraction of clean energy we will use, because the unit cost of clean power increases more slowly than that of brown power due to the smaller pollution factor $\gamma_{i,n}$, although the price $p_{i,n}$ of clean power is higher in general. However, when $p_{i,n}, \forall n \in \mathcal{N}_i$ are identical, the ratio of clean power is a constant and is independent of $Q_i^{buy}$. In this case, the numerator of $q_{i,n}^{*L}$ for $\forall n \in \mathcal{N}_i$ in Eq. (\ref{q-n solution}) are equal, thus the ratio $q_{i,1}^{*L}:\cdots:q_{i,N_i}^{*L}$ can be calculated as $\frac{1}{a_{i,1}}:\cdots:\frac{1}{a_{i,N_i}}$, which is constant. Furthermore, we can see that the fraction of clean power is sensitive to the configuration of the electricity price and the pollution factor $\gamma_{i,n}$. The fraction will be higher when the differences among $\gamma_{i,n}$ increase, or the differences among $p_{i,n}$ decrease, $\forall n \in \mathcal{N}_i$. 

In summary, we have verified the effects of PIF and the power factor $f^P$ in PIF. Our proposed PIF can encourage the power savings, the power buying from multiple sources and the usage of clean power. The more the power we buy, the stronger the effect on improving the usage of clean energy. In addition, the fraction of clean power is sensitive to several parameters, such as electricity price and pollution factor $\gamma_{i,n}$.

\subsection{Analysis of the Storage Scheduling}
\begin{figure} [!b]
	\centering
	\subfloat[$\eta^{'}_i=a_i*{\delta_i}+b_i$]{\includegraphics[width=4.1cm]{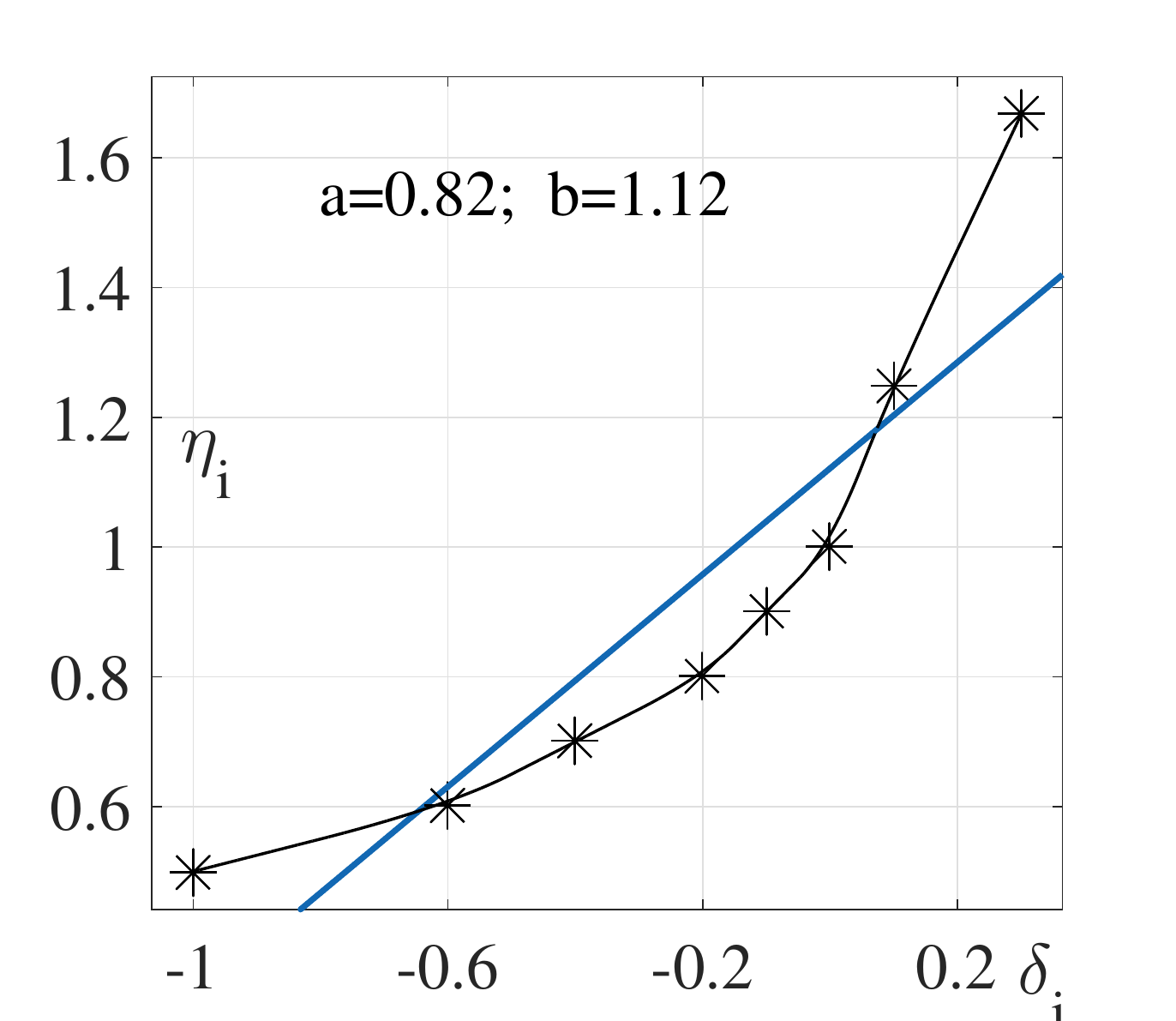}
							\label{niheyici}} 
						\hfill
	\subfloat[$\eta^{'}=a*\delta^2+b*\delta+c$]{\includegraphics[width=4.1cm]{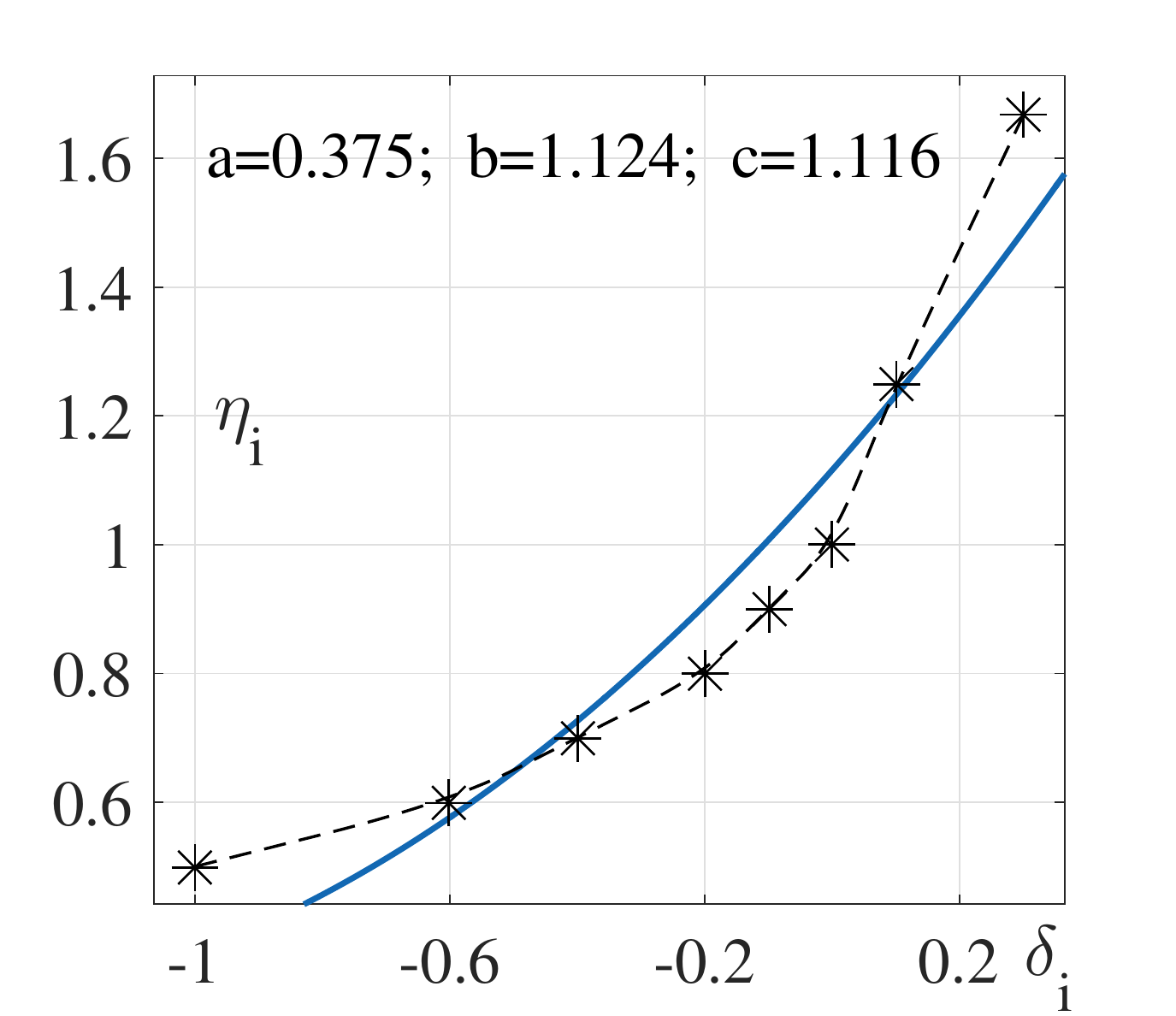}
		\label{niheerci2}} 
	\hfill
	\subfloat[$\eta^{'}=a*\delta^3+b*\delta^2+c*\delta+d$]{\includegraphics[width=4.1cm]{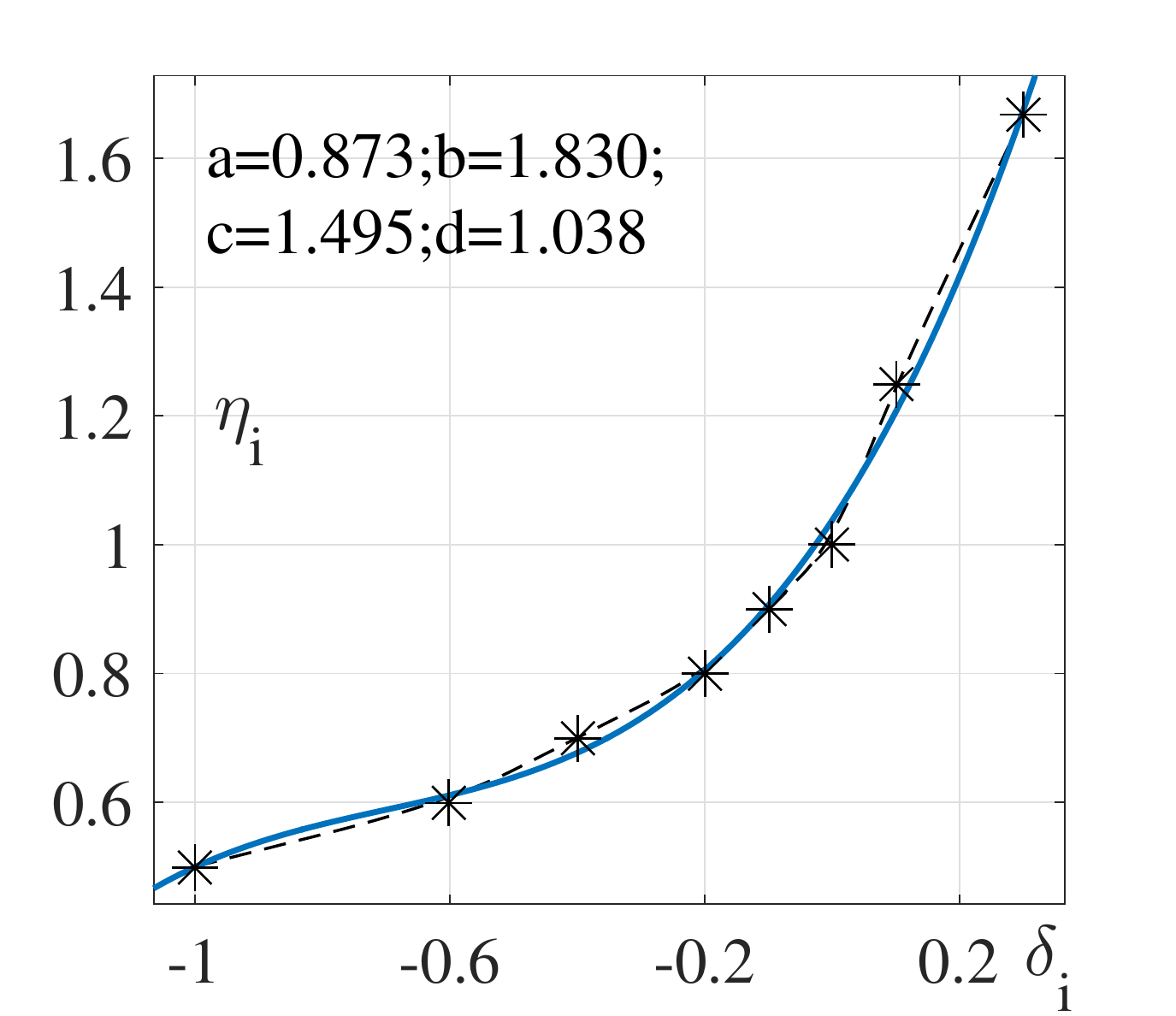}
		\label{nihesanci}} 
	\hfill
	\subfloat[$\eta^{'}=a*b^{\delta}+c$]{\includegraphics[width=4.1cm]{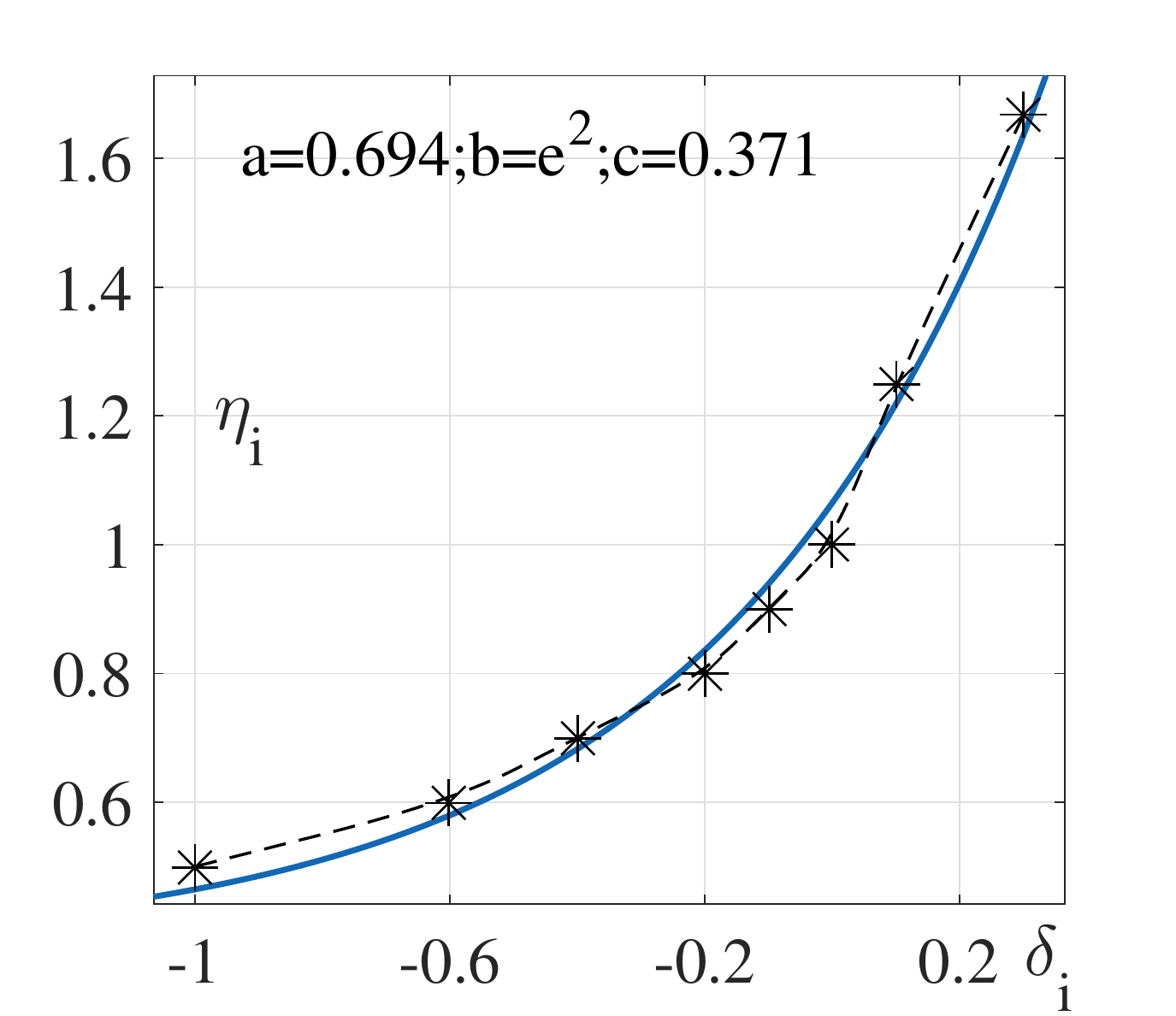}
		\label{nihezhishu}} 
	\caption{Performance of fitting $\eta^{'}$ with linear function, power function and exponential function}
	\label{nihetu} 
\end{figure}
First, we aim to give the parameters of charge and discharge efficiency $\eta^{'}$. Fig. \ref{nihetu} shows the results of fitting $\eta^{'}$ with exponential and power functions accounting for the convex constraints proved in Appendix B. Here, the black dotted lines represent the real values of $\eta^{'}$ and the blue solid lines represent the fitted values. It can be seen that the cubic function and the exponential function can cause smaller errors than the linear function and quadratic function. Besides, we find that setting $\eta^{'}$ to cubic function can converge about 1.5 to 2 times as fast as setting $\eta^{'}$ to exponential function when solving Problem P with Algorithm \ref{Algorithm3}. Therefore, in this paper, we model $\eta^{'}$ as a cubic function $0.873\delta^3+1.830\delta^2+1.495\delta+1.038$ where $\delta \in[-1,0.3]$, as is shown in Fig. \ref{nihesanci}.

\begin{figure}[!htbp]
	\centering
	\subfloat[Behaviors without potential cost]{\includegraphics[width=4.3cm]{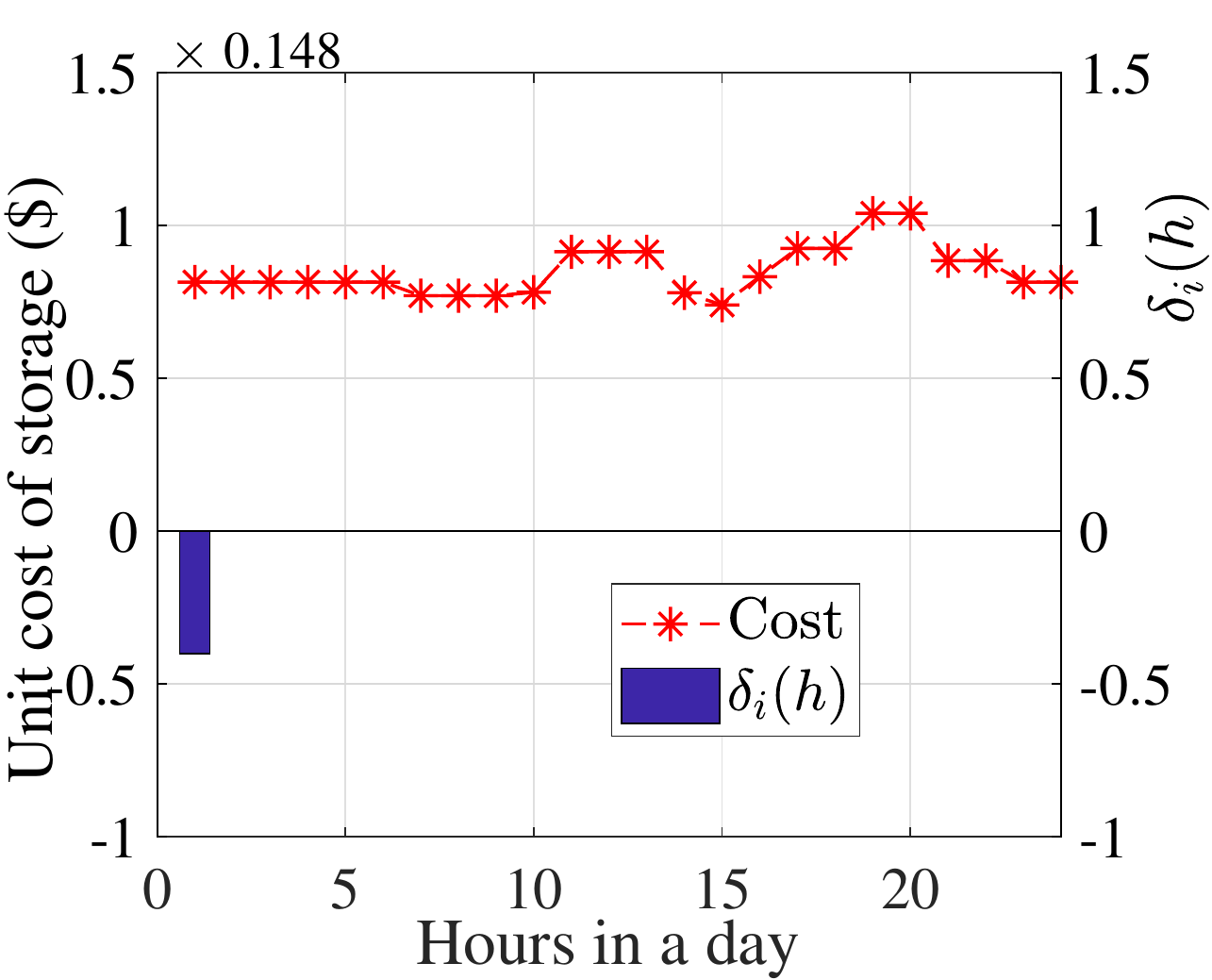}
		\label{WithoutPotentialCost}} 
	\hfill
	\subfloat[Behaviors with potential cost]{\includegraphics[width=4.3cm]{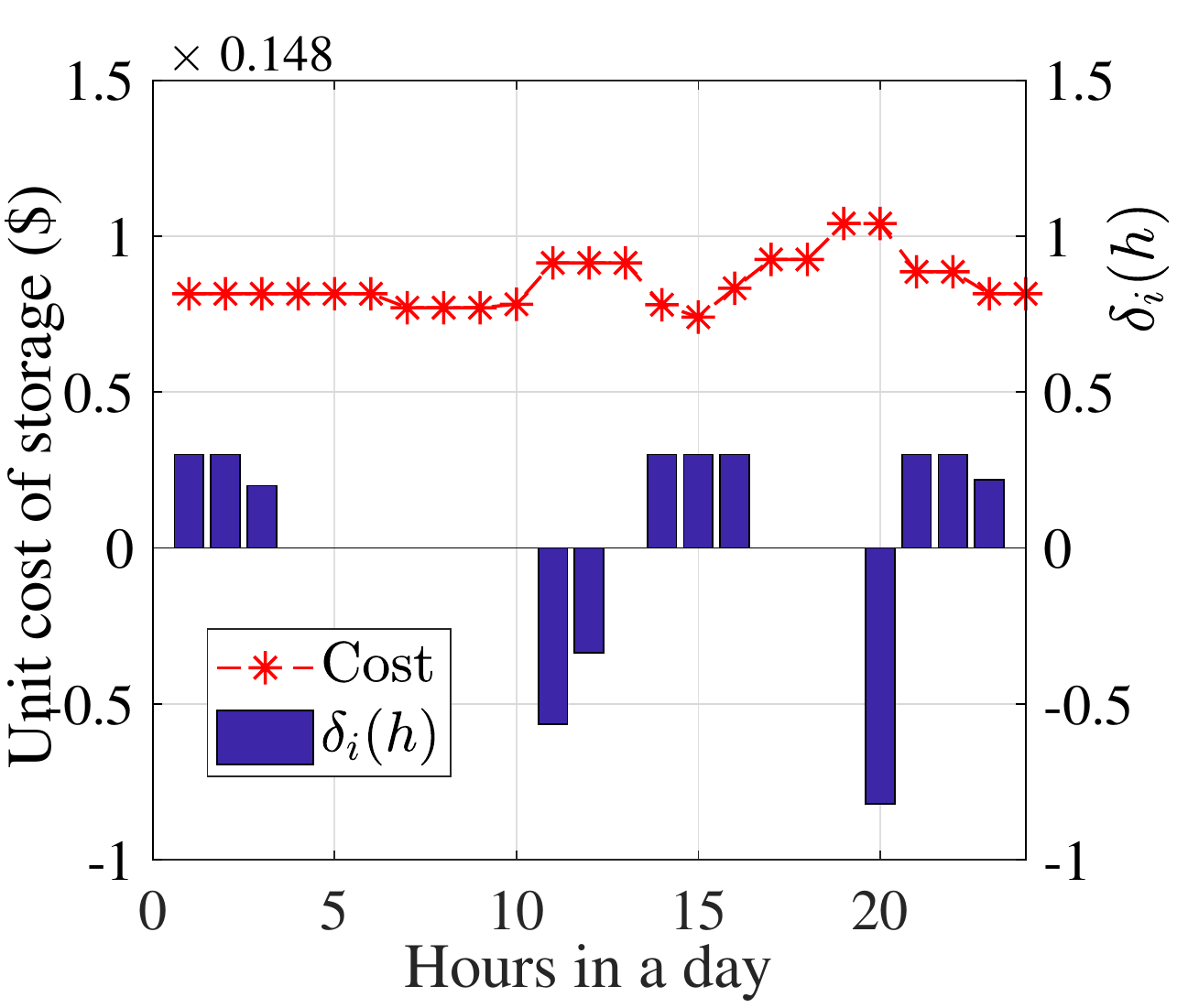}
		\label{WithPotentialCost}} 
		\caption{Impact of potential cost on storage scheduling and power cost.}
		\label{PotentialCost}	
\end{figure}

Secondly, we aim to verify the impact of potential cost on storage scheduling and power cost. Fig. \ref{WithoutPotentialCost} shows the optimal decisions on charge and discharge without considering potential cost, in which case batteries discharge much in the first time slot and do not charge or discharge in the remaining slots in order to obtain the maximum present interests while ignoring the future interests. Fig. \ref{WithPotentialCost} shows the optimal decisions on charge and discharge considering potential cost, where batteries always tend to discharge in slots with higher unit cost and charge in slots with lower unit cost, where $h$ is the indic of a future slot. This reveals that only by the model of potential cost, can we optimize the storage scheduling according to the varying unit-cost of storage in the long run.  

\begin{figure}
	\centering
	\subfloat[Potential cost $\hat{\varepsilon}_i$]{\includegraphics[width=4.4cm]{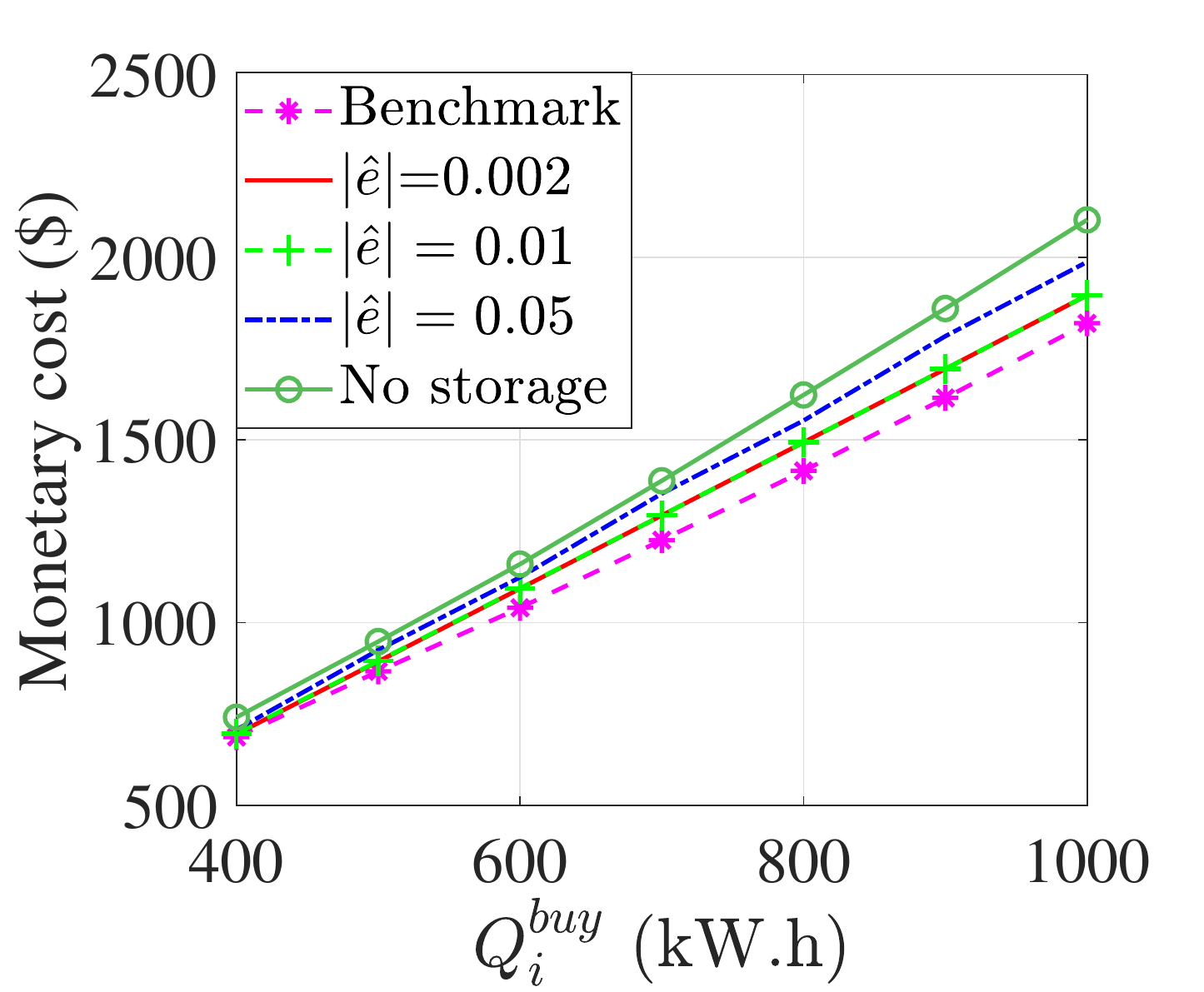}
		\label{epsilon1}} 
	\hfill
	\subfloat[Electricity price]{\includegraphics[width=4.2cm]{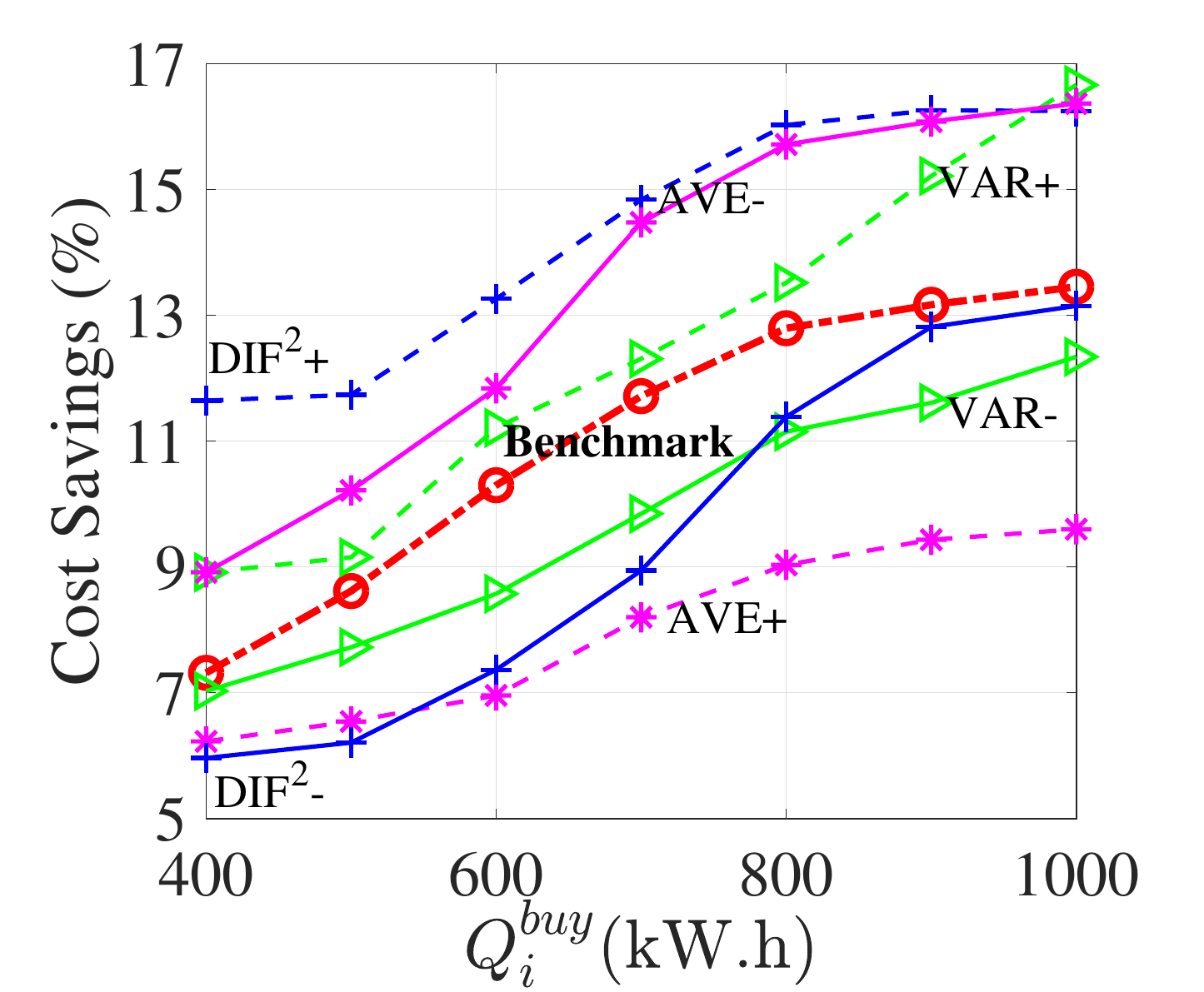}
		\label{Priceanalyse1}} 
	\caption{Sensitivity analysis of parameters for storage scheduling.}
\end{figure}

Fig. \ref{epsilon1} shows the daily monetary costs for a 1-MW data center with different accuracy of $\hat{\varepsilon}_i$ and that without $\hat{\varepsilon}_i$. It is observed that introducing potential cost can improve the power savings, but a small prediction error $\hat{e}$ of $\hat{\varepsilon}_i$ can cause much performance decline of storage scheduling. This means that the performance of storage scheduling is very sensitive to the parameter $\hat{\varepsilon}_i$, and an accurate prediction method of the unit cost of storage in the future slots is needed, which can be further studied in our future work. 

Thirdly, we aim to illustrate how the electricity price influences the power savings of storage scheduling, as is shown in Fig. \ref{Priceanalyse1}. We adopt the default electricity price as a benchmark, and change the average values 'AVE', the variance 'VAR', and the square sum of differences between prices in adjacent slots 'DIF', respectively, where '$+$' represents increase and '$-$' represents decrease. It is found that the ratios of monetary costs saved by storage scheduling will increase when 'VAR' or 'DIF' grows. This means that the higher the differences among electricity prices in different slots, the more the monetary costs saved by storage scheduling. However, the ratio of cost savings decreases when 'AVE' increases. Because when the electricity prices in different slots are equally increased, the differences among them do not change and the monetary costs with and without storage scheduling increase equally.

\subsection{Behavior of the Workload Scheduling}
\begin{figure}[!htbp]
	\centering
	\subfloat[Data center 1]{\includegraphics[width=4.3cm]{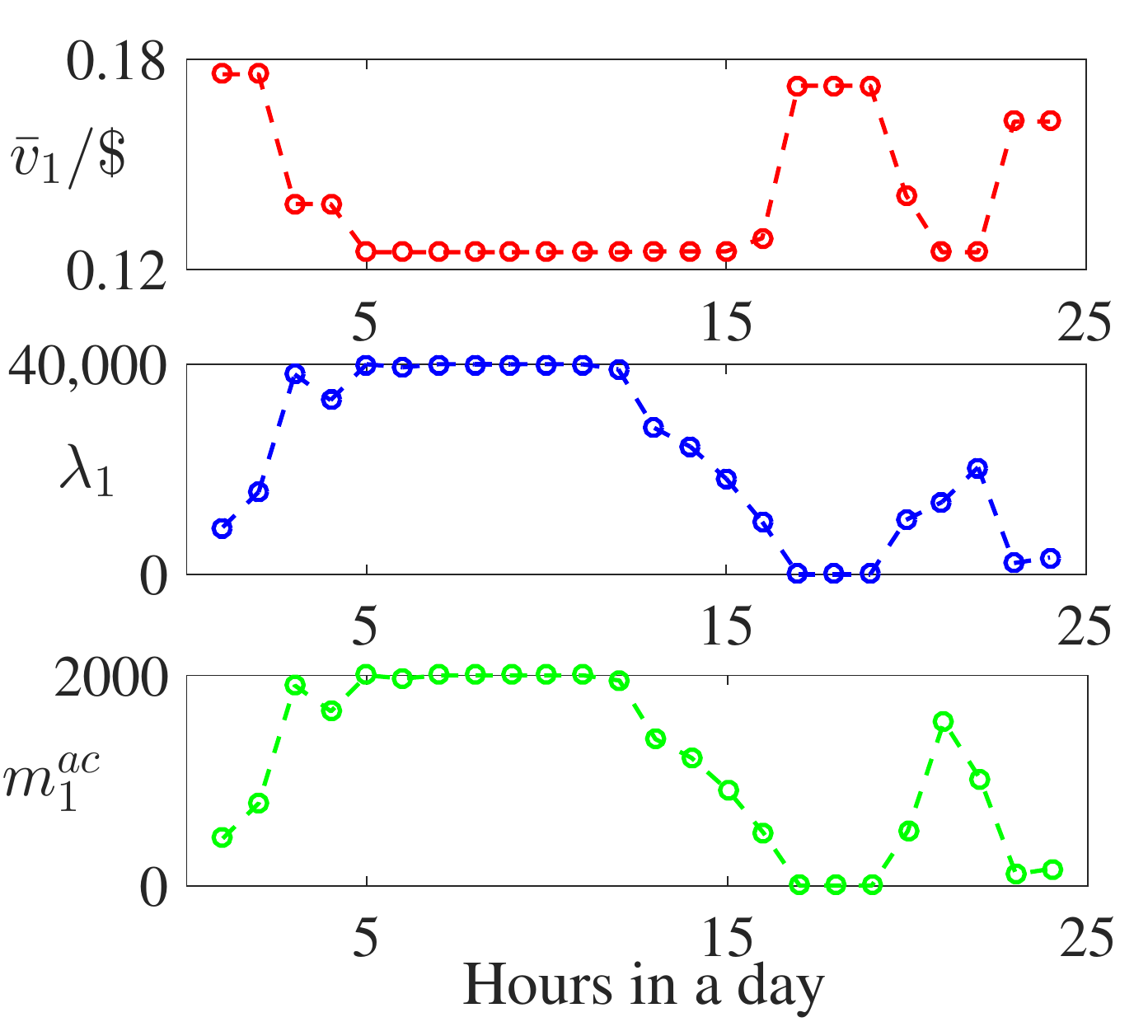}
		\label{Datacenter1}} 
	\hfill
	\subfloat[Data center 2]{\includegraphics[width=4.3cm]{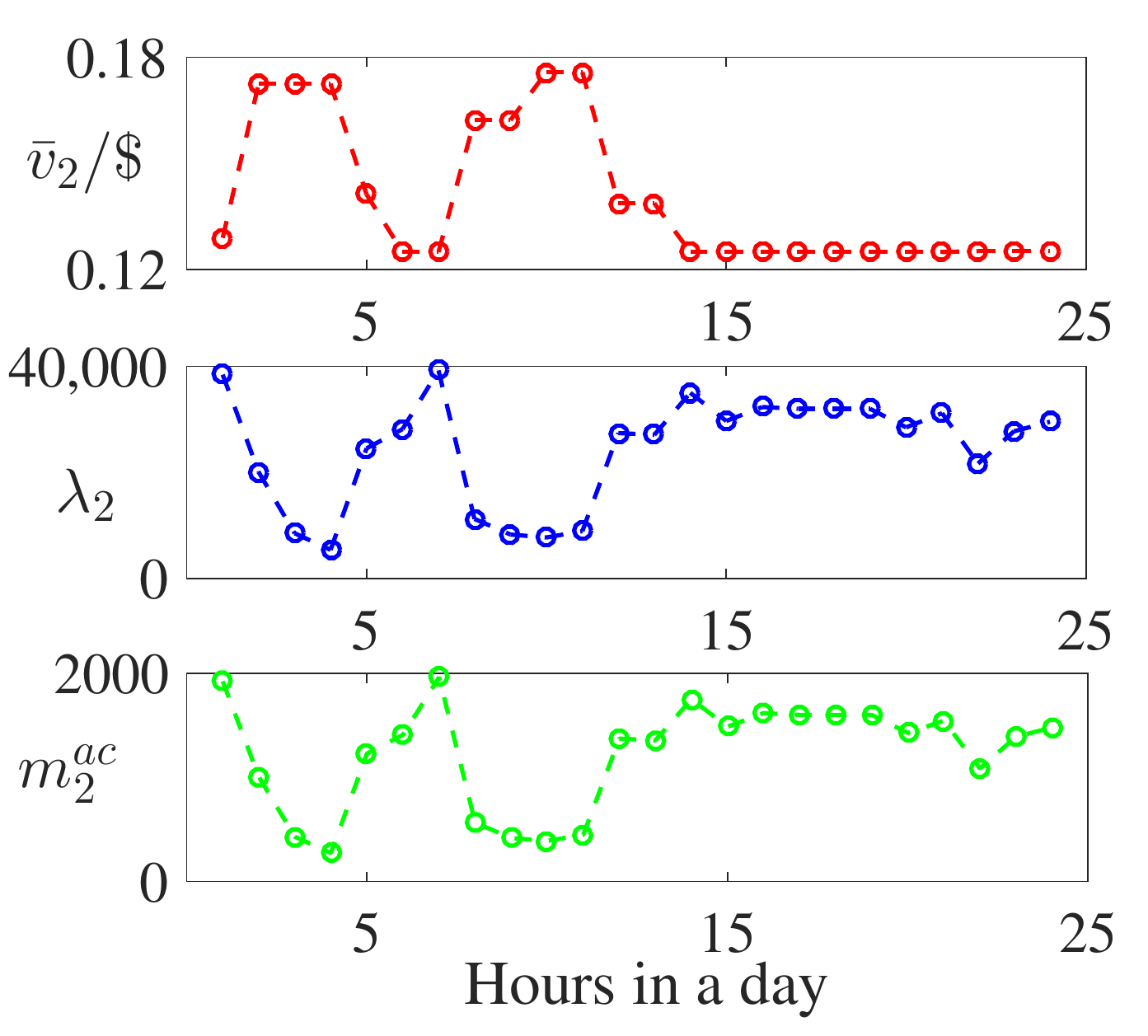}
		\label{Datacenter2}} 
	\hfill

	\caption{Diagram of daily $\bar{v}_i$, $m_i^{ac}$, $\lambda_i^{DC}$ of each data center}
	\label{miac}
\end{figure}

First, we assume that there are eight data centers and solve Problem P with Algorithm \ref{Algorithm3}. Fig. \ref{miac} shows the correlations among the unit cost of electricity $\bar{v}_i$, the arrival rate of requests $\lambda_i^{DC}$ and the number of active servers $m_i^{ac}$. It can be seen that $\lambda_i^{DC}$ increases with the decrease of $\bar{v}_i$, which means that more tasks will be dispatched to data centers with lower unit cost of electricity in order to reduce power costs. In addition, $m_i^{ac}$ decreases with the decrease of $\lambda_i^{DC}$, which means that the number of active servers can be flexibly adjusted according to the arrival rate of requests in order to save power energy. 

Secondly, Table \ref{workload} shows the monetary savings of the whole data center system in one time slot by workload scheduling with different configurations of electricity price $p_{i,n}$, the number of data centers $I$ and the total request rate $L$. It is observed that the larger the differences of electricity price, measured by the variances, in different areas, the more the monetary savings. However, when the prices in different areas increase equally, in which case only the means increase and the variances keep constant, monetary savings will not increase, because the differences among electricity prices are not changed. In addition, monetary savings will also increase with the growth of $I$ and $L$, which means that the effect of workload scheduling on power cost savings will be improved when the number of geographical data centers or the total requests rate increases.   
\begin{table}[htb] \small
	\centering
	\caption{Monetary costs saved by workload scheduling with different parameters.}
	\begin{tabular}{|l|c|c|c|c|c|}
		\hline
		\multicolumn{6}{|c|}{Monetary cost savings with different Electricity Price}\\ \hline
		Mean            & Standard            & \multicolumn{2}{c|}{Standard}         	   & Higher          &   Lower \\ \hline
		Variance        & Standard            & Higher         &   Lower                   & \multicolumn{2}{c|}{Standard}\\  \hline
		Savings(\$) 	& 436 	              & 967 	       & 176 	                   & 429 	         & 432   \\
		Percent         & 10.5\%              & 25.6\%		   & 4.1\%		               & 8.7\%		     & 12.9\% \\
		\hline
		\multicolumn{6}{|c|}{Monetary cost savings with the different number of data centers}\\ \hline
		$I$                   & 2            & 4           & 6          & 10           & 14  \\ \hline
		Savings(\$)  		  & 37           & 153         & 311        & 555          & 822   \\
		Percent               & 3.2\%        & 6.8\%       & 10.1\%		& 10.3\%	   & 11.2\%  \\
		\hline	
		\multicolumn{6}{|c|}{Monetary cost savings with different workload amount}\\ \hline
		$L/L_{max}$           & 5\%   		   & 20\%  			& 40\% 			& 60\% 			& 80\% \\ \hline
		Savings(\$)	 		  & 65			   & 186			& 340	  		& 436			& 323  \\
		Percent			  	  & 7.3\%		   & 10.8\%			& 11.8\%		& 10.5\%		& 5.9\% \\
		\hline	
	\end{tabular}
	\label{workload}
\end{table}

Thirdly, we can see from Table \ref{table1} that the mean values of $D^q_i$ are always limited between 0.1s and 0.2s, which is much less than the upper bound, and will not increase with the growth of $I$. This means that our formulation will never make the delay $D^q_i$ approach or exceed the upper bound that users can tolerate, which is a benefit for delay-sensitive requests and is different from the approaches presented in \cite{Shao2014Optimal}--\cite{Yao2014Power}. In addition, simulation results show that the variances of $D^q_i$ for $i\in \mathcal{I}$ are very tiny, which indicates that delay costs for different data centers are similar and our formulation can reduce the delay of requests while maintaining fairness.

\subsection{Evaluation of the Power Cost Reduction} 

First, Table \ref{powercost} shows the monetary cost savings obtained by the joint optimization of workload scheduling and storage scheduling in Problem P. It is observed that the monetary cost savings obtained by the joint optimization are larger than that obtained by workload scheduling in Table \ref{workload} or storage scheduling in Fig. \ref{Priceanalyse1} alone, and will increase with the growth of $I$, $L$ and the variance of electricity price. This means that $(\romannumeral 1)$ our proposed joint optimization model can further improve the power cost savings, and $(\romannumeral 2)$ the amount of cost savings is directly proportional to the number of data centers, the request rate, the spatial and the temporal change intensity of electricity price. Specifically, we find a ratio of cost savings of 10\%--30\%. 
\begin{table}[htb] \small
	\centering
	\caption{Monetary cost savings with workload scheduling and storage scheduling}
	\begin{tabular}{|l|c|c|c|c|c|c|}
		\hline
		\multirow{2}{*}{$L/L_{max}$}		& \multicolumn{3}{c|}{Standard Price} 		& \multicolumn{3}{c|}{Price with larger VAR} \\ \cline{2-7}
		& $I=4$		& $I=6$		& $I=8$		& $I=4$		& $I=6$		& $I=8$ \\	\cline{1-7}
		\multirow{2}{*}{40\%}	& 481	& 800	& 1027	& 592 	& 1089 	& 1390 \\ 
		& 15.1\%	& 17.0\%	& 17.5\%	& 21.1\%	& 26.1\%	& 25.2\%  \\  \cline{1-7}						
		\multirow{2}{*}{60\%}	& 708	& 1099	& 1529	& 798 	& 1323 	& 1745 \\ 
		& 14.5\%	& 15.0\%	& 15.7\%	& 18.2\%	& 20.4\%	& 20.1\% \\  \cline{1-7}
		\multirow{2}{*}{80\%}	& 930 	& 1454 	& 2016 	& 1280 	& 1941 	& 2598  \\
		& 13.8\%	& 14.4\%	& 14.9\%	& 20.7\%	& 21.1\%	& 21.2\% \\
		\hline												
	\end{tabular}
	\label{powercost}
\end{table}

\begin{figure}[htbp]
	\centering
	\includegraphics[width=4.7cm]{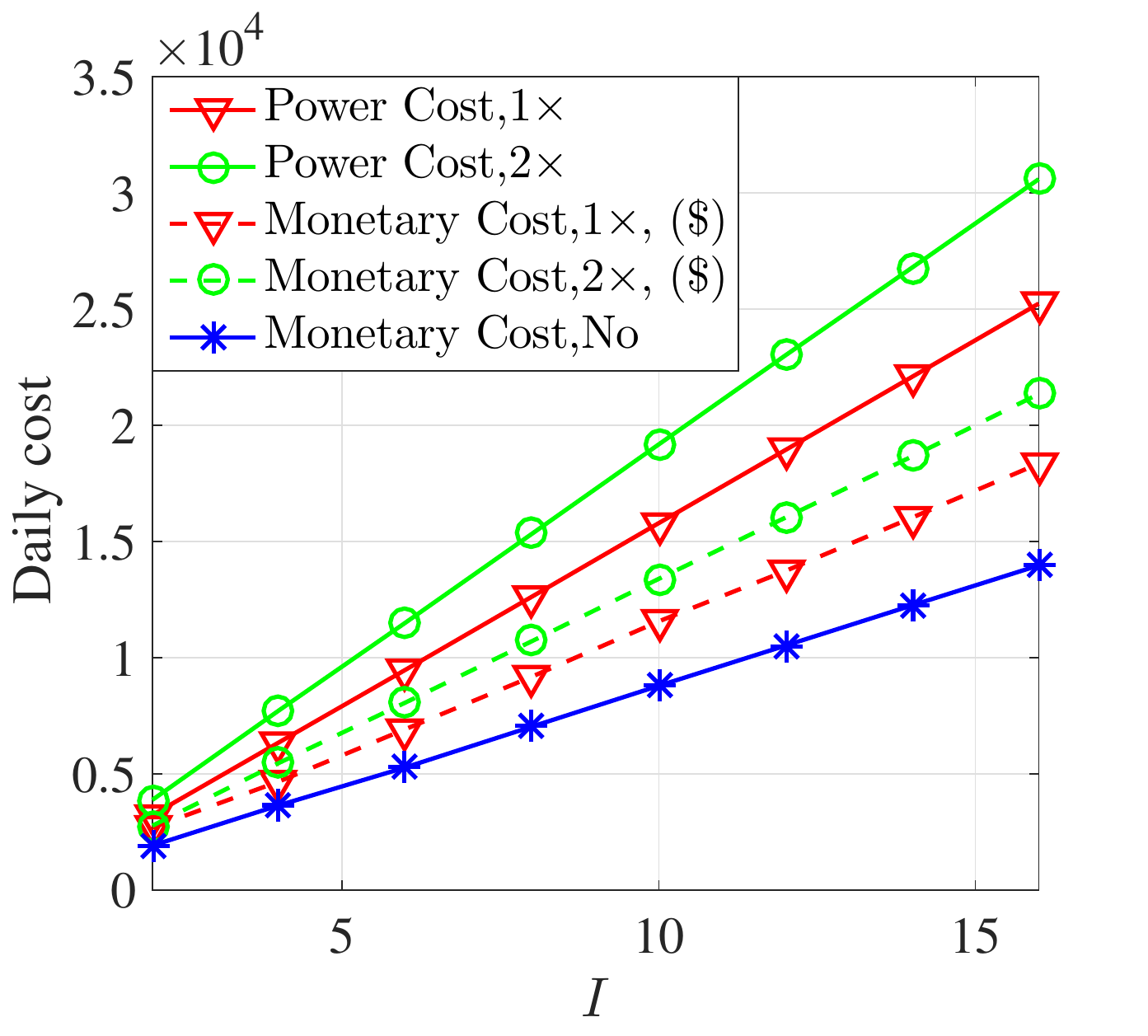} 
	\caption{The Power costs and monetary costs obtained by the joint optimization }
	\label{Powercost}
\end{figure}

Secondly, Fig. \ref{Powercost} shows the power costs and monetary costs with different pollution factors $\{\gamma_{i,n}\}$ in PIF and those without $PIF$, where '$1\times$' and '$2\times$' represent  $[\gamma_{i,TP},\gamma_{i,WP},\gamma_{i,SP}]$ equal to $[0.5,0.4,0.3]$ and $[1,0.8,0.6]$, respectively. The pollution costs measured by the pollution factors can be obtained by the differences of the power costs and the monetary costs. On the other hand, Table \ref{percent} shows the corresponding fractions of bought clean power. It is observed that when introducing the pollution factors or increasing the differences among the $\gamma_{i,n}$ of different power sources, the monetary costs grow, while the fraction of clean power is improved. This means that our proposed method can realize a trade-off between monetary costs and pollution costs, by which the fraction of clean power can be improved to 50\%--60\%, as is shown Table \ref{percent}.
		
\begin{table}
	\centering
	\caption{}
	\begin{tabular}{|l|c|c|c|}
		\hline
		$\gamma_{i,n}$		& TP		& WP		& SP	\\
		\hline
		No					& 81.34\%	& 18.66\%	& 0\%	\\
		$\times1$			& 53.63\%	& 36.68\%	& 9.68\%  \\
		$\times2$			& 39.49\%	& 42.73\%	& 17.78\%  \\
		\hline		
	\end{tabular}
	\label{percent}
\end{table}		

\section{Conclusion and future work} 
In this paper, we focused on the eco-friendly power cost minimization of geo-distributed data centers with multi-source power supply, where the schedulings of workload and power storage are both considered. We formulated this problem as Problem P\&W, and then simplified it to Problem P. To solve Problem P, we first proposed the SCP algorithm to obtain the globally optimal non-integer solution, and then proposed a low-complexity method, Algorithm \ref{Algorithm3}, to seek the quasi-optimal mixed-integer solution. Meanwhile, we proved the optimality of the SCP algorithm mathematically, and showed the performance of Algorithm \ref{Algorithm3} by simulation. Finally, Simulation results revealed that our method could effectively cut down the total power cost and encourage an eco-friendly use of power energy, as well as reduce the delay of requests, achieving energy cost savings of up to 10\%--30\%. More importantly, our proposed PIF-based power cost model can greatly improve the use of cleaner energy and encourage the savings of power. For example, for a 1-MW data center, the amount of clean energy it uses can reach 60\% of the total power it buys.  

As part of our future work, we will explore better methods to model the transmission delay and workload distribution.
\section*{Acknowledgement}
This work was supported by Overseas Expertise Introductions Center for Discipline Innovation ("111 Center") in China, and State Scholarship Fund awarded by the China Scholarship Council (No. 201806470032).
\section*{Appendix}
\begin{appendices}
	\subsection{Proofs about the optimality of Algorithm \ref{Algorithm1} and \ref{Algorithm2}}
	In this subsection, we aim to prove the optimality of Algorithm \ref{Algorithm1} and Algorithm \ref{Algorithm2}. In Proposition 1, we prove that the optimal solution of those $q_{i,n}$, which have $q_{i,n}^{*L}<0$, in problem (\ref{q-n}) is zero, which is also the optimal solution in each iteration of Algorithm \ref{Algorithm1} and Algorithm \ref{Algorithm2} according to inference. Thus in Proposition 2 and Proposition 3, we mainly give proofs of the final optimal solution of those $q_{i,n}$ which have $q_{i,n}^{*L}<0$. 
	\newtheorem{Pro11}{Proposition}
	\begin{Pro11}  
		For $\forall q_{i,n}$ in Problem (\ref{q-n}), if the optimal Lagrange solution $q_{i,n}^{*L}$ obtained by Eq. (\ref{q-n solution}) is less than zero, then its optimal solution $q_{i,n}^{*}$ of Problem (\ref{q-n}) with the constraint $q_{i,n}\geq0$ is zero, where $i \in \mathcal{I}, n \in \mathcal{N}_i$. 
	\end{Pro11}
	\begin{proof}
		First, we denote $q_{i,m}$ as the one whose optimal Lagrange solution $q_{i,m}^{*L}$ obtained by Eq. (\ref{q-n solution}) is less than zero, if there is any. Then it is true that $F^{PC}_i(\mathbf{q}_{i}^{*L})<F^{PC}_i(q_{i,m},\mathbf{q}_{i}^{*L}(q_{i,m}))$, where $\mathbf{q}_{i}^{*L}(q_{i,m})$ is the optimal Lagrange solution, obtained by Eq. (\ref{q-n solution}) after removing $m$ from $\mathcal{N}_i$, with the constraint $\sum\nolimits_{n\in\mathcal{N}_i \& n\ne m } q_{i,n} = Q_i^{cons} + Q_i^{bat}-q_{i,m}$.
			
		Secondly, we denote $f_n(q_{i,n})=a_{i,n}q_{i,n}^{2}+p_{i,n}q_{i,n}$, and rewrite (\ref{q-n}) as $F^{PC}_i(\mathbf{q}_{i})=\sum\nolimits_{n}a_{i,n}q_{i,n}^{2}+p_{i,n}q_{i,n}=\sum\nolimits_{n}f_n(q_{i,n})$. Then (\ref{B-3}) can be obtained as
			\begin{equation}\tag{A-1}\small
			\label{B-3}
			\begin{aligned}		
			\; & \qquad\qquad \qquad F^{PC}_i(\mathbf{q}_{i}^{*L})< F^{PC}_i(q_{i,m},\mathbf{q}_{i}^{*L}(q_{i,m}))& \Rightarrow \\
			\;-&\sum\nolimits_{n \ne m}\frac{\left\{f_n(q_{i,n}^{*L}(q_{i,m}))-f_n(q_{i,n}^{*L})\right\}}{q_{i,m}-q_{i,m}^{*L}}<\frac{f_m(q_{i,m})-f_m(q_{i,m}^{*L})}{q_{i,m}-q_{i,m}^{*L}}&\Rightarrow\\
			\; &\sum\nolimits_{n \ne m}\frac{\left\{f_n(q_{i,n}^{*L}(q_{i,m}))-f_n(q_{i,n}^{*L})\right\} }{\alpha_{i,n}\left\{q_{i,n}^{*L}(q_{i,m})-q_{i,n}^{*L}\right\}}< \frac{f_m(q_{i,m})-f_m(q_{i,m}^{*L})}{q_{i,m}-q_{i,m}^{*L}}&\Rightarrow\\
			\; &\sum\nolimits_{n \ne m}\frac{\left\{f_n(q_{i,n}^{*L}(0))-f_n(q_{i,n}^{*L})\right\} }{\alpha_{i,n}\left\{q_{i,n}^{*L}(0)-q_{i,n}^{*L}\right\}}\quad< \frac{f_m(0)-f_m(q_{i,m}^{*L})}{0-q_{i,m}^{*L}}&\quad.\\
			\end{aligned}
			\end{equation}
			Here, $q_{i,n}^{*L}(0)$ is the $q_{i,n}^{*L}$ obtained by Eq. (\ref{q-n solution}) when $q_{i,m}=0$, and $\alpha_{i,n}$ is equal to $a_{i,n}X_{i,m}$ according to (\ref{q-n solution}), where $X_{i,m}$ is the $X_i$ after eliminating $m$ from $\mathcal{N}_i$. 
			
			Thirdly, because all $f_n(\cdot)$ are strictly increasing and convex, and $q_{i,m}^{*L}<0<q_{i,m}^{+}$ where $q_{i,m}^{+}$ represents $q_{i,m}>0$, we can obtain (\ref{B-4}) as follows.
			\begin{equation}\tag{A-2}\small
			\label{B-4}
			\begin{aligned}		
			\frac{f_m(q_{i,m}^{+})-f_m(0)}{q_{i,m}^{+}-0}>\frac{\mathrm{d}f_m(x) }{\mathrm{d} x} \bigg{|}_{x=0}>\frac{f_m(0)-f_m(q_{i,m}^{*L})}{0-q_{i,m}^{*L}}
			\end{aligned}
			\end{equation}
			In addition, it can be proved by contradiction that $q_{i,n}^{*L}(q_{i,m}^{+})<q_{i,n}^{*L}(0)<q_{i,n}^{*L}$. Thus we can obtain (\ref{B-5}) as below.
			\begin{equation}\tag{A-3}\small
			\label{B-5}
			\begin{aligned}		
			\frac{f_n(q_{i,n}^{*L}(q_{i,m}^{+}))-f_n(q_{i,n}^{*L}(0))}{q_{i,n}^{*L}(q_{i,m}^{+})-q_{i,n}^{*L}(0)}<\frac{\mathrm{d}f_n }{\mathrm{d} x} \bigg{|}_{x=q_{i,n}^{*L}(0)}<\frac{f_n(q_{i,n}^{*L}(0))-f_n(q_{i,n}^{*L})}{q_{i,n}^{*L}(0)-q_{i,n}^{*L}}
			\end{aligned}
			\end{equation} 
			
			Finally, based on(\ref{B-3}), (\ref{B-4}) and (\ref{B-5}), we can obtain (\ref{B-6}) as
			\begin{equation}\tag{A-4}\small
			\label{B-6}
			\begin{aligned}	
			&\sum\nolimits_{n \ne m}\frac{\left\{f_n(q_{i,n}^{*L}(q_{i,m}^{+}))-f_n(q_{i,n}^{*L}(0))\right\} }{\alpha_{i,n}\left\{q_{i,n}^{*L}(q_{i,m}^{+})-q_{i,n}^{*L}(0)\right\}}< \frac{f_m(q_{i,m}^{+})-f_m(0) }{q_{i,m}^{+}-0}&\Rightarrow\\	
			&\sum\nolimits_{n\ne m}\left\{f_n(q_{i,n}^{*L}(q_{i,m}^{+}))-f_n(q_{i,n}^{*L}(0))\right\}\,<f_m(q_{i,m}^{+})-f_m(0)&\Rightarrow\\
			& \qquad\qquad
			F^{PC}_i(0,\mathbf{q}_{i}^{*L}(0))<F^{PC}_i(q_{i,m}^{+},\mathbf{q}_{i}^{*L}(q_{i,m}^{+}))\, .
			\end{aligned}
			\end{equation}
		This means that the optimal solution of $q_{i,m}$ in (\ref{q-n}) whose $q_{i,m}^{*L}<0$ with the constraint $q_{i,m}\geq0$ is zero. Based on the above, the case that there is only one $q_{i,n}^{L*}<0$ has been proved. Then we can similarly prove the case with two, three or more $q_{i,n}^{L*}<0$.
	\end{proof}	
	\begin{Pro11}
		Algorithm \ref{Algorithm1} can obtain the optimal solution of Problem (\ref{q-n}) for the $i$th data center.
	\end{Pro11}
	\begin{proof}
		First, in the $t$-th, $t\in\{1,\cdots,T\}$ iteration of Algorithm \ref{Algorithm1}, where T is the total number of iterations, the optimal solution of any $q_{i,n}$ which has $q_{i,n}^{L*}<0$ is equal to zero according to Proposition 1. 
		
		Secondly, for $\forall q_{i,n}, n\in \mathcal{N}_i$ whose $q_{i,n}^{L*}\geq0$, the constraint $q_{i,n}\geq0$ is a slack constraint\cite{boyd2004convex}, so that its optimal solution in (\ref{q-n}) is the $q_{i,n}^{L*}$ obtained by Eq.{\ref{q-n solution}}.
		
		Thirdly, we need to prove that zero is the final optimal solution of the $q_{i,n}$ set to zero in the $t$-th iteration, although it is optimal in the $t$-th iteration. When $t=T-1$, the optimal solutions of the remaining $q_{i,n}$ can be obtained in the $T$th iteration, all of which are no less than zero, and there are no changes of the $q_{i,n}$ set to zero in the $(T-1)$-th iteration. When $1\leq t<T-1$, there is at least one new $n$ added into $\mathcal{N}_i^{-}$ in the $(t+1)$-th iteration. We denote $q_{i,n_1}$ as the one set to zero in the $t$-th iteration, and then denote $q_{i,n_1}^{*L,t}$ and $\nu_i^{*L,t}$ as the optimal Lagrange solution of $q_{i,n_1}$ and the corresponding marginal cost obtained by (\ref{XY}) before setting $q_{i,n_1}$ to zero in the $t$-th iteration. Considering $q_{i,n}^{*L}(q_{i,m}^{+})<q_{i,n}^{*L}(0)<q_{i,n}^{*L}$ as stated in the proof of Proposition 1, when we set $q_{i,n_1}=0$ in the $t$-th iteration, the optimal Lagrange solutions of the remaining $q_{i,n}$ obtained in the $(t+1)$-th iteration are all less than those obtained in the $t$-th iteration. Thus $\nu_i^{*L,t}>\nu_i^{*L,t+1}$, which can be further extended to $\nu_i^{*L,t_1}>\nu_i^{*L,t_2}, t_1<t_2$. The marginal cost of $q_{i,n_1}^{*L,t}$ is $\nu_i^{*L,t}$. If $q_{i,n_1}=0$ is not the finally optimal solution, then it would be recalculated as in (\ref{q-n solution}) in the $(t+m)$-th iteration, $m=\{2,\cdots,T-t-1\}$, which must be larger than zero, and the corresponding marginal cost is $\nu_i^{*L,t+m}$. There is $\nu_i^{*L,t+m}>\nu_i^{*L,t}$, because $q_{i,n_1}^{*L,t+m}>0>q_{i,n_1}^{*L,t}$, which is a contradiction. Therefore, we can conclude that zero is the final optimal solution of the $q_{i,n}$ set to zero in the $t$-th iteration.
		
		Based on the above, the complete $\mathcal{N}_i^{-}$ can be obtained in the $(T-1)$-th iteration, in which the optimal solutions of all $q_{i,n}$ are zero, and the optimal solutions of all $q_{i,n}\in\widetilde{\mathcal{N}}_i^{-}$ can be solved in the $T$-th iteration. In addition, it can be easily proved that $T\leq \max \{N_1,\cdots,N_I\}+1$, so that it is true that Algorithm \ref{Algorithm1} can always converge and can obtain the optimal solution of Problem (\ref{q-n}) for the $i$th data center.
	\end{proof}
		\begin{Pro11}
			Algorithm \ref{Algorithm2} is guaranteed to obtain the optimal non-integer solution of Problem P$^*$ shown in (\ref{Cost1})
		\end{Pro11}
		\begin{proof}
			It is noted that we regard Problem P1$^*$ as a convex problem, which will be proved in Appendix B. We see that whether the optimal $q_{i,n}\geq0$ is obtained influences the values of the constants $X_i$, $Y_i$ and $Z_i$ in (\ref{Cost1-equal}a). When the optimal $q_{i,n}$ limited by $q_{i,n}\geq0$ are all obtained, Problem P$^*$ is equal to Problem P1$^*$ and can be solved by SQP. Then we will prove that all optimal $q_{i,n}\geq0$ can be obtained by Algorithm \ref{Algorithm2} as follows.		

			Different from Algorithm \ref{Algorithm1}, the final optimal solution of the $q_{i,n}$ set to zero in the $t$-th iteration may be larger than zero. Thus we add Lines 10--12 into Algorithm \ref{Algorithm2} to move the $n$ whose $q_{i,n}^{L*,t}<0$ but $q_{i,n}^{*}>0$ back to $\widetilde{\mathcal{N}}_i^{-}$, which is the main difference between Algorithm \ref{Algorithm1} and Algorithm \ref{Algorithm2}. We denote $q_{i,m}$ as the one whose $q_{i,n}^{L*,t}<0$ but $q_{i,n}^{*}>0$, denote $\epsilon^{+}>0$ as an infinitesimal positive number, and denote $F_i^{PC,t+m}$ as the optimal power cost for the $i$-th data center in the $(t+m)$-th iteration, where $m=\{1,\cdots,T-t\}$. The marginal cost of $q_{i,m}=0$ is $p_{i,m}$ and those of all $q_{i,n}$ solved by Lagrange method in the $(t+m)$-th iteration are $\nu_i^{*L,t+m}$. When keeping the total power amount $\sum\limits_{n\in\mathcal{N}_i\&n\ne m}q_{i,n}+q_{i,m}$ unchanged, $\mathop {\lim }\limits_{\epsilon^{+}\to0}F_i^{PC,t+m}(\epsilon^{+})= F_i^{PC,t+m}(0)+p_i^m\cdot\epsilon^{+}-\nu_i^{*L,t+m}\cdot\epsilon^{+}$, where $F_i^{PC,t+m}(*)$ represents the optimal $F_i^{PC,t+m}$ when $q_{i,m}=*$. If $p_{i,m}<\nu_i^{*L,t+m}$, then $F_i^{PC,t+m}(0)>F_i^{PC,t+m}(\epsilon^{+})$, which means $q_{i,m}=0$ is not the optimum and $m$ needs to be moved back to $\widetilde{\mathcal{N}}_i^{-}$. Otherwise, $F_i^{PC,t+m}(0)<F_i^{PC,t+m}(\epsilon^{+})$, which means $q_{i,m}=0$ is still the optimum in the $(t+m)$-th iteration, and is the final optimum when $m=T-t$. 
			
			In addition, it can be easily inferred that $T\leq2*(\sum\limits_{i=1}^{I}N_i)$. There are two steps leading to the iterations in SCP algorithm, including $(\romannumeral1)$ setting those $q_{i,n}=0$ whose optimal Lagrange solution $q_{i,n}^{*L}<0$ temporarily, $(\romannumeral2)$ recalculating their optimal values if zeros are not the optimal solutions. For the worst case (but never happening), all $q_{i,n}^{*L}$ are less than zero, which causes $\sum\limits_{i=1}^{I}N_i$ iterations. Meanwhile, those final optimal solutions are all more than zero instead of equal to zero, which causes another $\sum\limits_{i=1}^{I}N_i$ iterations, so that the algorithm can converge with no more than $2*(\sum\limits_{i=1}^{I}N_i)$ iterations.
			
			In conclusion, Algorithm \ref{Algorithm2} is guaranteed to obtain the optimal solutions of all $q_{i,n}$ limited by $q_{i,n}\geq0$, so that it is sure to obtain the optimal non-integer solution of Problem P$^*$.						
	
		\end{proof}	
	\subsection{Proofs about convexity conditions of Problem P1$^*$}
	
	In this section, we will give proofs about conditions that $\tilde\eta_i'(\delta_i)$ needs to meet to guarantee the convexity of Problem P1$^*$ defined on the convex set $\mathbb{C}$. Here, $\mathbb{C}$ is a convex set composed of all constraints in (\ref{Cost1-equal}). 
	\begin{Pro11}
		For any type of $\eta_i^{'}(\delta_i)$ whose first and second partial derivatives are all existent and continuous, as long as it meets the condition that $\small{2}\cdot\frac{\mathrm{d}\eta_i^{'} }{\mathrm{d} \Delta_i}+ \frac{\mathrm{d^2}\eta_i^{'} }{\mathrm{d} (\Delta_i)^2}\cdot \small{\Delta_i} \geq 0$ on the convex set $\mathbb{C}$, where $\forall i\in \mathcal{I}$ and $\delta_i=\Delta_i/C_i$, then  
		Problem P1$^*$ is a convex programming problem.
	\end{Pro11}
	\begin{proof}[Proof]
		For any type of $\eta_i^{'}(\delta_i)$ whose first and second partial derivatives are all existent and continuous, $\forall i\in \mathcal{I}$, we can obtain the first and second partial derivatives of $\Phi_i$ as follows.
		\begin{flalign*}\small
			& \frac{\partial{\Phi_i}}{\partial {\lambda _i^{DC}}}=\theta_{1,i}
			\left\{\frac{1}{(m_i^{ac}\bar{u_i}-{\lambda _i^{DC}})^2}\right\}&\\
			& \frac{\partial{\Phi _i}}{\partial m_i^{ac}}=\theta_{1,i}\left\{\frac{-\bar{ u}_i}{(m_i^{ac}\bar{u}_i-\lambda_i^{DC})^2}\right\}  
			+\theta_{2,i} \cdot s_i^{\alpha} \left\{ \frac{2(P_i^{cons} + \eta_i^{'}\Delta_i)}{X_i} +{Y_i} \right\}&\\
			& \frac{\partial \Phi_i}{\partial \Delta _i} = \theta_{2,i}\Bigg{\{} \frac{2(P_i^{cons} + \eta_i^{'} \Delta _i)+X_iY_i}{X_i}\cdot \frac{\mathrm{d}(\eta_i^{'}\Delta_i) }{\mathrm{d} \Delta_i} - \hat \varepsilon_i \Bigg{\}}&\\ 
			& \frac{\partial ^2 \Phi _i}{\partial (\lambda_i^{DC})^2} = \theta _{1,i} \left\{ \frac{2}{(m_i^{ac}\bar{u}_i-\lambda_i^{DC})^3} \right\} &\\
			& \frac{\partial^2 \Phi _i}{\partial (m_i^{ac})^2} = \theta _{1,i} \left\{ \frac{2(\bar{u}_i)^2}{(m_i^{ac}\bar {u}_i-\lambda_i^{DC})^3} \right\} 
			+ \theta_{2,i} \left\{ \frac{2(s_i^{\alpha})^2}{X_i} \right\} &\\
			& \frac{\partial^2 \Phi _i}{\partial (\Delta _i)^2} =\frac{\theta _{2,i}}{X_i} \Bigg{\{} 2(\frac{\mathrm{d}(\eta_i^{'}\Delta_i) }{\mathrm{d} \Delta_i})^2 + \bigg{\{}2P_i^{cons}+2\eta_i^{'} \Delta _i+X_iY_i \bigg{\}}\frac{\mathrm{d^2}(\eta_i^{'}\Delta_i) }{\mathrm{d} (\Delta_i)^2}\Bigg{\}} &\\
			& \frac{\partial^2\Phi _i}{\partial \lambda_i^{DC} m_i^{ac}} =\frac{\partial ^2 \Phi _i}{\partial m_i^{ac}\lambda_i^{DC}} = \theta _{1,i}\left\{ \frac{ -2\bar {u}_i}{(m_i^{ac}\bar{u}_i - \lambda_i^{DC})^3} \right\}&\\
			& \frac{\partial ^2 \Phi _i}{\partial \lambda_i^{DC}\Delta _i} = \frac{\partial ^2\Phi _i}{\partial \Delta _i\lambda_i^{DC}} = 0&\\
			& \frac{\partial ^2\Phi _i}{\partial m_i^{ac}\Delta _i} = \frac{\partial ^2\Phi _i}{\partial \Delta _im_i^{ac}} = \frac{2\theta _{2,i}s_i^{\alpha}}{X_i} \left\{ \frac{\mathrm{d}(\eta_i^{'}\Delta_i) }{\mathrm{d} \Delta_i} \right\}&\\
		\end{flalign*} 
		Based on this, the Hessian Matrix $\Lambda_i^{DC}$ of $\Phi_i$ in Problem P1$^*$ can be easily obtained, and the Leading Principal Minors $\Lambda_{i,1}$, $\Lambda_{i,2}$ and $\Lambda_{i,3}$ of $\Lambda_i^{DC}$ are given as follows.

		\begin{flalign*}\small
			& \Lambda_{i,1}=  \frac{2 \theta _{1,i}}{(m_i^{ac}\bar{u}_i-\lambda_i^{DC})^3} &\\
			& \Lambda_{i,2}=\frac{4\theta _{1,i} \theta_{2,i}(s_i^{\alpha})^2}{X_i (m_i^{ac}\bar{u}_i-\lambda _i^{DC})^3} & \\   
			& \Lambda_{i,3}=\frac{8\theta _{1,i}\theta _{2,i}^2 (s_i^{\alpha})^2 (P_i^{cons} + \eta _i^{'}\Delta _i + 0.5X_iY_i)}{(X_i)^2(m_i^{ac}\bar{u}_i -\lambda_i^{DC})^3} \cdot\frac{\mathrm{d^2}(\eta_i^{'}\Delta_i) }{\mathrm{d} (\Delta_i)^2}&                   
		\end{flalign*} 
		As we have stated that $m_i^{ac}\bar{u}_i-\lambda_i^{DC}>0 $, $Q_i^{cons} + \eta _i^{'}\Delta _i\geq 0$, $X_i>0$, $Y_i>0$, etc. above, it can be concluded that both $\Lambda_{i,1}>0$ and $\Lambda_{i,2}>0$ are true on the convex set $\mathbb{C}$. $\Lambda_{i,3} \geq 0$ is also true in the case of $\frac{\mathrm{d^2}(\eta_i^{'}\Delta_i) }{\mathrm{d} (\Delta_i)^2}=\small{2}\cdot\frac{\mathrm{d}\eta_i^{'} }{\mathrm{d} \Delta_i}+ \frac{\mathrm{d^2}\eta_i^{'} }{\mathrm{d} (\Delta_i)^2}\cdot \small{\Delta_i}\geq 0$. Therefore, if $\small{2}\cdot\frac{\mathrm{d}\eta_i^{'} }{\mathrm{d} \Delta_i}+ \frac{\mathrm{d^2}\eta_i^{'} }{\mathrm{d} (\Delta_i)^2}\cdot \small{\Delta_i} \geq 0$, the  Hessian Matrix of $\Phi_i$ is positive semidefinite on convex set $\mathbb{C}$. 
		
		Furthermore, because a sum of convex functions is also a convex function according to \cite{boyd2004convex}, $\Phi$ is convex on the convex set $\mathbb{C}$ and Problem P1$^*$ is a CP problem when $\small{2}\cdot\frac{\mathrm{d}\eta_i^{'} }{\mathrm{d} \Delta_i}+ \frac{\mathrm{d^2}\eta_i^{'} }{\mathrm{d} (\Delta_i)^2}\cdot \small{\Delta_i} \geq 0\, ,\, \forall i\in \mathcal{I}$ for any $\eta_i^{'}(\delta_i)$ whose first and second partial derivatives are all existent and continuous.

	\end{proof}

\end{appendices}
\bibliographystyle{unsrt}  
\bibliography{References}

\begin{IEEEbiography}[{\includegraphics[width=1in,height=1.25in,clip,keepaspectratio]{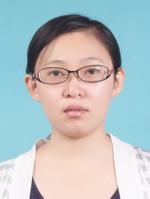}}]{Chunlei Sun}
	
	received the Bachelor's degree in Shandong University in 2014. And now she is pursing the PH.D degree in Beijing University of Posts and Telecommunications. Her researches include demand side management in energy internet, energy efficiency optimization of data center, etc.  Email: scl1992@bupt.edu.cn.
\end{IEEEbiography}	
\begin{IEEEbiography}[{\includegraphics[width=1in,height=1.25in,clip,keepaspectratio]{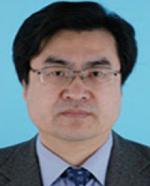}}]{Xiangming Wen}
	is the director of Beijing Key Laboratory of Network System Architecture and Convergence, where he has managed several projects related to open wireless networking. He is also the vice president of Beijing University of Posts and Telecommunications. He received both his M.Sc. and Ph.D in information and communication engineering from Beijing University of Posts and Telecommunications. His current research interests focus on radio resource and mobility management, software defined wireless networks, and broadband multimedia transmission technology. Email: xiangmw@bupt.edu.cn
\end{IEEEbiography}	
\begin{IEEEbiography}[{\includegraphics[width=1in,height=1.25in,clip,keepaspectratio]{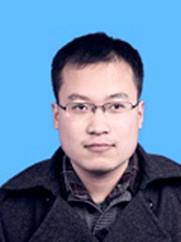}}]{Zhaoming Lu}
	received the Ph.D in Beijing University of Posts and Telecommunications in 2012. He joined the School of Information and Communication Engineering in Beijing University of Posts and Telecommunications in 2012. His researches include Open Wireless Networks, QoE management in wireless networks, software defined wireless networks, cross-layer design for mobile video applications and so on. Email: lzy0372@bupt.edu.cn	
\end{IEEEbiography}	
\begin{IEEEbiography}[{\includegraphics[width=1in,height=1.25in,clip,keepaspectratio]{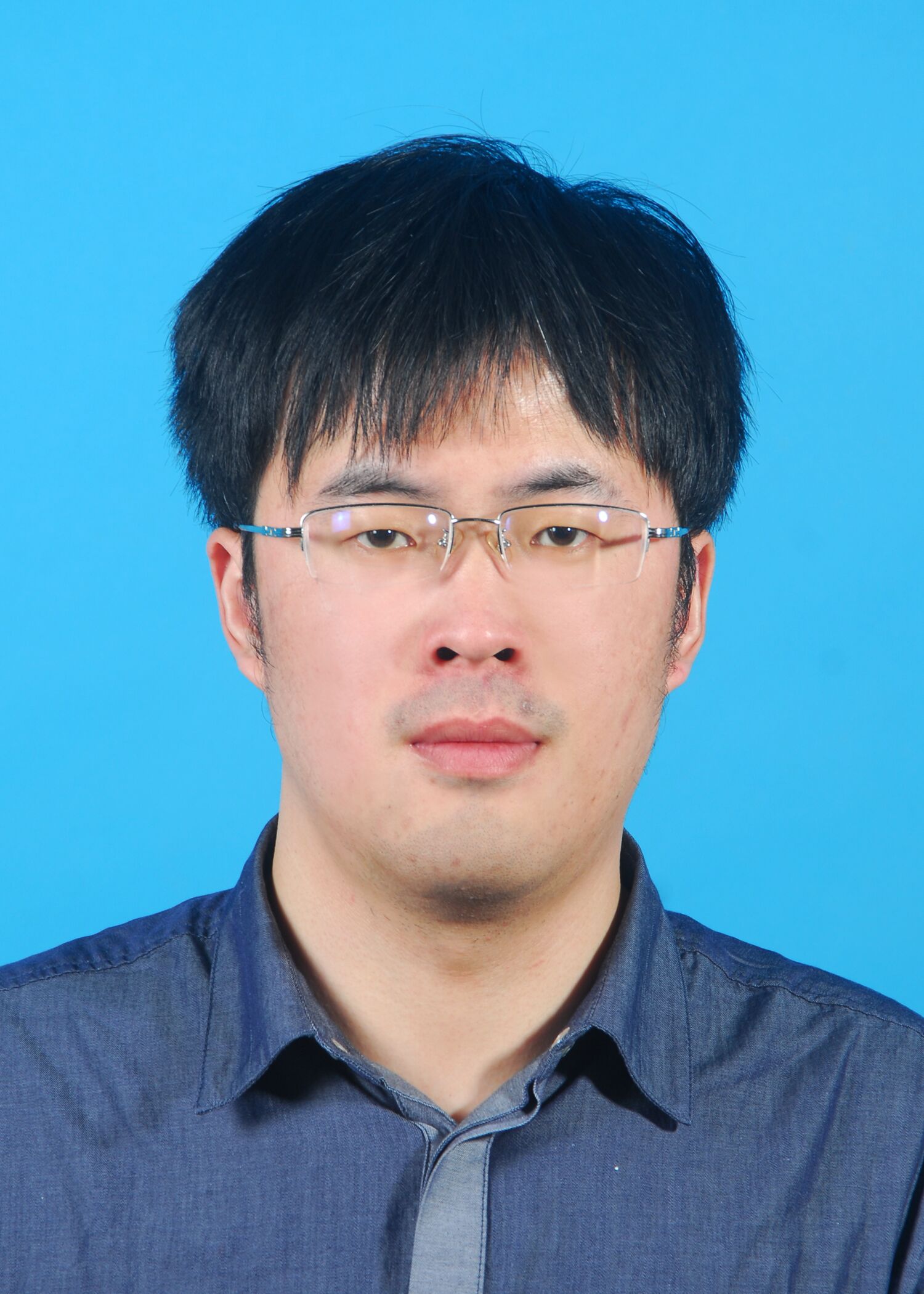}}]{Wenpeng Jing}
	received his B.S.degree in Communication Engineering from Shandong University in 2012 and the Ph.D degree from Beijing University of Posts and Telecommunications in 2017. He is currently a postdoc in Beijing University of Posts and Telecommunications. His research interests include green communication, radio resource allocation, energy efficiency optimization in heterogeneous networks.	Email: jingwenpeng@bupt.edu.cn
\end{IEEEbiography}	
\begin{IEEEbiography}[{\includegraphics[width=1in,height=1.25in,clip,keepaspectratio]{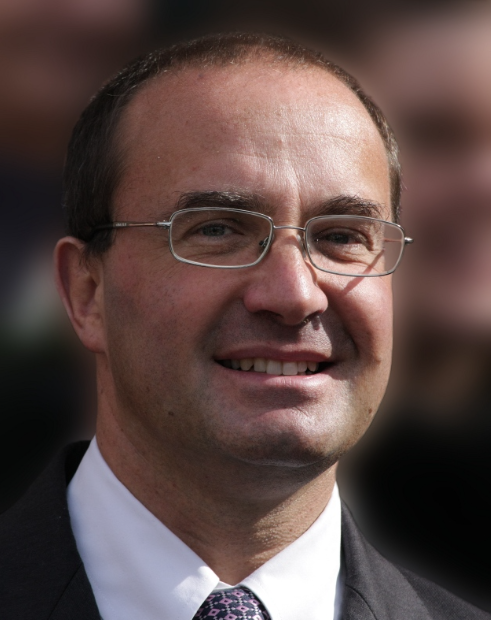}}]{Michele Zorzi}(F'07) received the Laurea and Ph.D.degrees in electrical engineering from the University of Padova in 1990 and 1994, respectively. From 1992 to 1993, he was on leave with the University of California at San Diego (l JCSD). After being affiliated with the Dipartimento di Elettron-ica e Informazione, Politecnico di Milano, Italy, the Center for Wireless Communications, UCSD, and the University of Ferrara, in 2003, he joined the Faculty of the Information Engineering Department,University of Padova, where he is currently a Professor. His current research interests include performance evaluation in mobile communications systems, random access in mobile radio networks, ad hoc and sensor networks, energy-constrained communications protocols, 5G millimeter-wave cellular systems, and underwater communications and networking. He was an Editor-in-Chief of the IEEE Wireless Communications from 2003 to 2005 and an Editor in-Chief of the IEEE Transactions on Communications from 2008 to 2011. He was a Guest Editor for several special issues in the IEEE Personal Communications, the IEEE Wireless Communications, the IEEE Network, and the IEEE Journal on Selected Areas in Communications. He served as a Member at-Large in the Board of Governors of the IEEE Communications Society from 2009 to 2011,and as its Director of Education from 2014 to 2015. He is currently the Founding Editor-in-Chief of the IEEE Transactions on Cognitive Communications and Networks. Email: zorzi@dei.unipd.it
	
\end{IEEEbiography}

\end{document}